\documentclass{IEEEtran}

\usepackage[utf8]{inputenc}
\usepackage[T1]{fontenc}
\usepackage{amsmath, amssymb, amsthm}
\usepackage{booktabs}
\usepackage{multirow}
\usepackage{xspace}
\usepackage{hyperref}
\usepackage[all]{hypcap}
\usepackage{graphicx}
\usepackage{subfig}
\usepackage{listings}
\usepackage[table]{xcolor}
\usepackage{tikz}
\usepackage{pgfplots}
\usepackage{lipsum}

\colorlet{graybg}{gray!10}
\colorlet{plot1}{red}
\colorlet{plot2}{green!75!black}
\colorlet{plot3}{blue}
\colorlet{plot4}{yellow!75!gray}
\colorlet{plot5}{magenta}
\colorlet{plot6}{cyan}
\colorlet{plot7}{orange}
\colorlet{plot8}{brown}

\colorlet{plot1bg}{plot1!20}
\colorlet{plot2bg}{plot2!20}
\colorlet{plot3bg}{plot3!20}
\colorlet{plot4bg}{plot4!20}

\colorlet{todo}{blue!50!black}
\colorlet{ready}{black}

\usetikzlibrary{
    external,
    scopes,
    patterns,
    decorations.pathreplacing,
    calc,
    positioning,
    plotmarks,
    fadings,
    3d
}
\pgfplotsset{
    every axis/.append style={
        axis background/.style={
            fill=graybg
        },
        xmin=0,
        ymin=0,
        xlabel near ticks,
        ylabel near ticks,
        enlarge x limits={
            value=0.05,
            auto
        },
        enlarge y limits={
            value=0.05,
            auto
        },
    },
    twocolplot/.style={
        width=.26\textwidth,
    },
    onecolplot/.style={
        width=.45\textwidth,
    },
}
\pgfkeys{
    /tikz/external/mode=list and make
}
\tikzsetexternalprefix{figures/externalized/}
\tikzset{external/export=false}

\hypersetup{
    pdftitle={Performance Modeling for Dense Linear Algebra},    
    pdfauthor={Elmar Peise and Paolo Bientinesi},     
    pdfsubject={},   
    pdfkeywords={blocked algorithms} {performance modelling} {ranking}, 
    pdfnewwindow=true,      
    colorlinks=true,       
    linkcolor=black,          
    citecolor=black,        
    filecolor={.5 0 0},      
    urlcolor=black           
}

\newcommand{\appendixref}[1]{\hyperref[#1]{Appendix~\ref*{#1}}}

\numberwithin{figure}{section}

\renewcommand{\mathtt}[1]{\text{\ttfamily #1}}
\newcommand{\overparameq}[2]{%
    \tikz[baseline=(n.base)] {
        \node[inner sep=0pt, label=90:] (n) {\texttt{#2\vphantom{gT$A$}}};
        \node[inner sep=0pt, anchor=south, above=1pt of n.north] (l)
            {\scriptsize \textcolor{gray}{\texttt{#1\vphantom{gT}}}};
        \node at (l.north) {};
    }%
}

\newcommand{\tikzsquare}[1]{\tikz \filldraw[fill=#1] (0, 0) rectangle (.25, .25);}

\usepackage[OMLmathsfit, sfdefault=iwona, scaled=1.0]{isomath}
\newcommand{\metric}[1]{\ensuremath{\text{\small\itshape\sffamily #1}}}

\renewcommand{\metric}[1]{\ensuremath{\mathsfit{#1}}}

\title{Performance Modeling for Dense Linear Algebra}

\author{
    Elmar Peise and Paolo Bientinesi\\
    \small AICES, RWTH Aachen\\
    Schinkelstr. 2\\
    52062 Aachen, Germany\\
    \texttt{\{peise,pauldj\}@aices.rwth-aachen.de}
}

\newcommand{\diff}[1]{#1}

\begin{document}
    
    \maketitle

    \begin{abstract}

    It is well known that the behavior of dense linear algebra algorithms is greatly influenced by factors like target architecture, underlying libraries and even problem size; because of this, the accurate prediction of their performance is a real challenge.
    In this article, we are not interested in creating accurate models for a given algorithm, but in correctly ranking a set of equivalent algorithms according to their performance.
    Aware of the hierarchical structure of dense linear algebra routines, we approach the problem by developing a framework for the automatic generation of statistical performance models for BLAS and LAPACK libraries.
    This allows us to obtain predictions through evaluating and combining such models.
    We demonstrate that our approach is successful in both single- and multi-core environments, not only in the ranking of algorithms but also in tuning their parameters.

\end{abstract}

    \includegraphics[height=0pt]{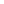}

    \section{Introduction}
    \label{sec:intro}
    For a large class of dense linear algebra operations, such as the solution of
least squares problems and linear systems, not one but many algorithms exist.
While mathematically they are equivalent, their performance depends ---in
different ways--- on factors such as the target architecture, the 
underlying
libraries, and the problem size. The prediction of the best performing
algorithm in a given scenario is a challenging task. Our goal is to \emph{rank
the algorithms} according to their performance and to determine their optimal
configuration \emph{without executing them}.

As a motivating example, we consider the inversion of a lower triangular matrix ($L \leftarrow
L^{-1}$), for which there exist four blocked algorithms, all
equivalent in exact arithmetic, but with different performance signatures. Each
such algorithmic variant depends on one parameter, the block-size $b$, which
determines the stride in which the matrix is traversed.

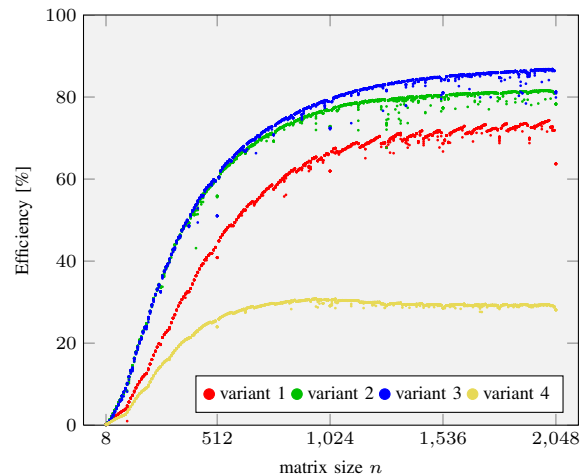
\begin{figure}[t]
    \scriptsize
    \centering
    \tikzset{external/export=true}

    \begin{tikzpicture}
        \begin{axis}[
            onecolplot,
            xlabel={matrix size $n$},
            ylabel={Efficiency [\%]},
            ymax=100,
            xmin={},
            xtick={8,512,1024,...,2048},
            legend columns=-1,
            legend pos=south east
        ]
            \addlegendimage{plot1, only marks}
            \label{fig:intro.motivation.trinv:v1}
            \addlegendentry{variant 1}
            \addlegendimage{plot2, only marks}
            \label{fig:intro.motivation.trinv:v2}
            \addlegendentry{variant 2}
            \addlegendimage{plot3, only marks}
            \label{fig:intro.motivation.trinv:v3}
            \addlegendentry{variant 3}
            \addlegendimage{plot4, only marks}
            \label{fig:intro.motivation.trinv:v4}
            \addlegendentry{variant 4}

            \addplot[plot1, mark size=.3pt, only marks] file {figures/data/intro.motivation.trinv/trinv1.n.eff.dat};
            \addplot[plot2, mark size=.3pt, only marks] file {figures/data/intro.motivation.trinv/trinv2.n.eff.dat};
            \addplot[plot3, mark size=.3pt, only marks] file {figures/data/intro.motivation.trinv/trinv3.n.eff.dat};
            \addplot[plot4, mark size=.3pt, only marks] file {figures/data/intro.motivation.trinv/trinv4.n.eff.dat};
        \end{axis}
    \end{tikzpicture}

    \caption{Inversion of a lower triangular matrix: Efficiency as a function of the problem size.}
    \label{fig:intro.motivation.trinv:en}
    \tikzset{external/export=false}
\end{figure}

\begin{figure}[t]
    \scriptsize
    \centering
    \tikzset{external/export=true}

    \begin{tikzpicture}
        \begin{axis}[
            onecolplot,
            xlabel={block-size $b$},
            ylabel={Efficiency [\%]},
            ymax=100,
            xmin=0,
            enlarge x limits=true,
            xtick={8,64,128,...,256},
            legend columns=-1,
            legend pos=south east
        ]
            \addlegendimage{plot1, only marks}
            \label{fig:intro.motivation.trinv:v1}
            \addlegendentry{variant 1}
            \addlegendimage{plot2, only marks}
            \label{fig:intro.motivation.trinv:v2}
            \addlegendentry{variant 2}
            \addlegendimage{plot3, only marks}
            \label{fig:intro.motivation.trinv:v3}
            \addlegendentry{variant 3}
            \addlegendimage{plot4, only marks}
            \label{fig:intro.motivation.trinv:v4}
            \addlegendentry{variant 4}

            \addplot[plot1, mark size=.3pt, only marks, restrict x to domain=0:256] file {figures/data/intro.motivation.trinv/trinv1.b.eff.dat};
            \addplot[plot2, mark size=.3pt, only marks, restrict x to domain=0:256] file {figures/data/intro.motivation.trinv/trinv2.b.eff.dat};
            \addplot[plot3, mark size=.3pt, only marks, restrict x to domain=0:256] file {figures/data/intro.motivation.trinv/trinv3.b.eff.dat};
            \addplot[plot4, mark size=.3pt, only marks, restrict x to domain=0:256] file {figures/data/intro.motivation.trinv/trinv4.b.eff.dat};
        \end{axis}
    \end{tikzpicture}

    \caption{Inversion of a lower triangular matrix: Efficiency as a function of the block-size.}
    \label{fig:intro.motivation.trinv:eb}
    \tikzset{external/export=false}
\end{figure}

In~\autoref{fig:intro.motivation.trinv:en} we plot their efficiency 
---the relative measure of the machines resource utilization---
when executed on one core of an Intel Harpertown E5450;
\footnote{
    The algorithms were implemented in C, compiled with ICC version 12.0, and
    linked to Intel's MKL version 10.2.6.
    \diff{In all cases, our algorithms use library calls to perform calculations;
    because of this, compiler's optimizations do not influence the resulting performance.}
} 
the block-size is fixed to $96$ and the matrix size $n$ varies.
The results show noticeable
differences in performance between algorithms:
Variant~4~(\ref{fig:intro.motivation.trinv:v4}) is significantly slower than
the others, while, especially for large matrices,
variant~3~(\ref{fig:intro.motivation.trinv:v3}) is the most efficient. In
\autoref{fig:intro.motivation.trinv:eb}, we fix $n = 1000$ and let $b$ vary;
in all variants, the efficiency decreases for small and large block-sizes. For
variants 1~(\ref{fig:intro.motivation.trinv:v1}),
2~(\ref{fig:intro.motivation.trinv:v2}), and
3~(\ref{fig:intro.motivation.trinv:v3}) the optimal choice of $b$ is close to
$100$.

This example shows that in order to reach high efficiency, it is
crucial to both single out the right algorithmic variant, and optimize
the block-size.  Due to the complexity of the architecture and the
memory access patterns, it is virtually impossible to perform these
tasks only by analyzing the mathematics of the algorithms. Indeed, 
experience tells us that 
the best choice heavily depends on the computational kernels
used, such as BLAS, on the processor architecture, and the matrix
size; changing any of these factors may lead to entirely different
performance behavior.

In this article, we detail a strategy based on the analysis of the BLAS
routines upon which the target algorithms are built; 
we introduce a tool that, using measurements, creates performance models
for BLAS kernels and stores them permanently in a repository. 
When faced with a set of algorithms, the models are
evaluated and combined to predict the algorithms' performance. These
predictions allow us not only to accurately rank the algorithmic variants,
but also to determine the optimal algorithmic block-size.

Several different approaches to performance modeling in dense linear algebra
exist; some notable examples are given in the following.
Cuenca et al.~developed
a system of self-optimizing linear algebra routines (SOLAR)~\cite{solar};
every routine is associated with performance information, which is
hierarchically propagated to higher level routines in order to tune them.
Dongarra et al.~proposed an approach for 
parallel software such as HPL and ScaLAPACK~\cite{hplmodeling};
they employ sampling and polynomial fitting to construct models in order
to extrapolate the performance of routines for larger problems and higher
parallelism.
Dackland et al.~predict the performance of ScaLAPACK algorithms through
models based on the efficiency of BLAS and the time spent on communication~\cite{scalapckanalysis}.
Balaprakash et al.~apply mathematical optimization techniques to reduce the
number of measurements in empirical performance tuning~\cite{kobe}.
Iakymchuk et al.~model the performance of BLAS analytically based
on memory access patterns~\cite{roman}; while their models represent the
program execution very accurately, constructing them requires a high level of
expertise of both routines and architecture.

In contrast to the aforementioned approaches, we aim at the automatic
generation of accurate models for BLAS routines, which constitute the building
blocks of a multitude of algorithms in linear algebra. Our main goal is not to obtain
accurate prediction for these algorithms, but rather to correctly rank them and
tune their configuration.

This article is structured as follows. In \autoref{sec:sampling}, we discuss
the performance of dense linear algebra routines. In \autoref{sec:modeling}, we
introduce the Modeler, a tool that automatically generates analytical
performance models. Predictions and ranking are discussed in
\autoref{sec:ranking}, and in \autoref{sec:conclusion}
we draw conclusions.

    \section{Performance}
    \label{sec:sampling}
    In this section, we discuss the concept of performance in the context
of dense linear algebra, and introduce {\em the Sampler}, a
performance measurement tool for linear algebra routines.

        \subsection{Performance Metrics}
        \label{sec:sampling.goal.metrics}
        In the following, the term performance is used broadly to cover 
a set of \emph{performance metrics} that
describe certain aspects of a routine execution, such as timings,
instruction counts, and cache accesses. The metrics are either directly 
obtained from hardware performance counters or are quantities computed from them.
The most fundamental performance counter ---the time stamp counter--- is provided by
a register that is incremented once per CPU cycle. It is accessed
through the x86 instruction \texttt{RDTSC} and serves 
as a cycle-accurate timer; we refer to this metric as \metric{ticks}.
In order to access more CPU performance counters,
we use the Performance Application Programming Interface (PAPI)~\cite{papi}\footnote{PAPI version 4.2.1.0.}.
PAPI provides
functions to configure, initialize, and read up to 107 counters, 
but usually only a subset of which are available in a given system.

In this article we focus on the highly accurate time metric \metric{ticks}. In additions, we use the derived metric
\metric{efficiency}, representing the relative resource utilization:
$$
    \metric{efficiency} = \frac{\metric{flops}}{\metric{ticks} \cdot \metric{fips}}.
$$
This measures how efficiently an operation that performs \metric{flops}
floating point instructions\footnote{%
    A fused multiply add operation $a \leftarrow b + c \cdot d$ is counted
    as a single floating point operation, since it is one instruction and
    processed as such by the CPU.
} uses the CPU's ALUs, which can perform up to
\metric{fips} floating point instructions per cycle.

        \subsection{Performance of dense linear algebra routines}
        \label{sec:sampling.goal.input}
        At this point, we are interested in the performance of dense linear
algebra routines, such as BLAS or unblocked algorithms, that act as
building blocks for higher level algorithms. Our first objective is to
build performance models for such building blocks.

For a given architecture, we regard the performance of a routine as a 
function of the arguments. Apart from the buffers for matrices and vectors,
all the arguments are simple to represent in such a function, since they are basic data types such as characters, integers, and floating
point numbers.
Since the instructions performed by
the dense linear algebra routines we consider are mostly independent of the input data, 
we can reduce the information needed for these
arguments to their size and storage location in the memory hierarchy.

Regarding  memory locality, we distinguish two cases: in-cache and out-of-cache.
\emph{In-cache} refers to the situation where all matrices are as close to the
CPU as possible, that is, in the lowest cache level that can accommodate
them. Since the access time is minimized, this scenario leads to the
best performance the routine can attain. \emph{Out-of-cache} 
refers to the opposite situation, where the matrices reside 
in main memory, thus causing costly data
transfers. Since the loading and storing of data might result in memory
stalls, the overall performance is often inferior than when data resides in cache.

\diff{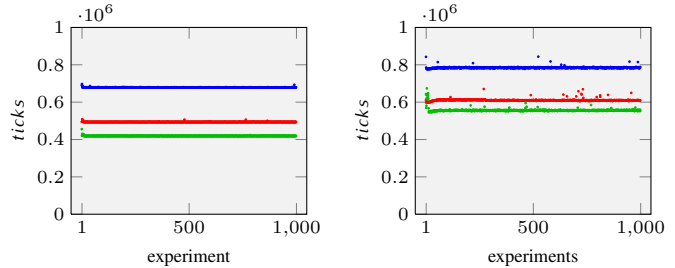
\begin{figure}[t]
    \scriptsize
    \centering
    \tikzset{external/export=true}

    \ref*{fig:sampling.experiments.ref:legend}

    \vspace{.1cm}

    \begin{tikzpicture}
        \begin{axis}[
            twocolplot,
            xlabel={experiment},
            ylabel={\metric{ticks}},
            ymax=1e6,
            xmin={},
            xtick={1, 500, 1000},
            legend to name=fig:sampling.experiments.ref:legend,
            legend columns=-1,
        ]
            \addlegendimage{plot1, only marks}
            \label{fig:sampling.experiments.ref.plot:open}
            \addlegendentry{OpenBLAS}
            \addlegendimage{plot2, only marks}
            \label{fig:sampling.experiments.ref.plot:mkl}
            \addlegendentry{MKL}
            \addlegendimage{plot3, only marks}
            \label{fig:sampling.experiments.ref.plot:atlas}
            \addlegendentry{ATLAS}

            \addplot[color=plot1, only marks, mark size=.3pt] table[x expr=\coordindex, y index=10] {figures/data/sampling.experiments/open.repeat.ic.dat};
            \addplot[color=plot2, only marks, mark size=.3pt] table[x expr=\coordindex, y index=10] {figures/data/sampling.experiments/mkl.repeat.ic.dat};
            \addplot[color=plot3, only marks, mark size=.3pt] table[x expr=\coordindex, y index=10] {figures/data/sampling.experiments/atlas.repeat.ic.dat};
        \end{axis}
    \end{tikzpicture}
    \hfill
    \begin{tikzpicture}
        \begin{axis}[
            twocolplot,
            xlabel={experiments},
            ylabel={\metric{ticks}},
            ymax=1e6,
            xmin={},
            xtick={1, 500, 1000}
        ]
            \addplot[color=plot1, only marks, mark size=.3pt] table[x expr=\coordindex, y index=10] {figures/data/sampling.experiments/open.repeat.oc.dat};
            \addplot[color=plot2, only marks, mark size=.3pt] table[x expr=\coordindex, y index=10] {figures/data/sampling.experiments/mkl.repeat.oc.dat};
            \addplot[color=plot3, only marks, mark size=.3pt] table[x expr=\coordindex, y index=10] {figures/data/sampling.experiments/atlas.repeat.oc.dat};
        \end{axis}
    \end{tikzpicture}

    \caption{Repeated execution of \texttt{dtrsm}: In-cache (left) and out-of-cache (right) operands.}
    \label{fig:sampling.experiments.plot}
    \tikzset{external/export=false}
    \scriptsize
\end{figure}}

To study the influence of memory locality and the reproducibility of measurements,
let us consider a repeated execution of the BLAS routine \texttt{dtrsm} ($B \leftarrow
A^{-1} B$, $A$ triangular). The interface is
\begin{center}
    \small
    \texttt{%
        dtrsm(%
        \overparameq{side}{R},
        \overparameq{uplo}{L},
        \overparameq{transA}{N},
        \overparameq{diag}{U},
        \overparameq{m}{512},
        \overparameq{n}{128},
        \\
        \overparameq{alpha}{0.37},
        \overparameq{A}{$A$},
        \overparameq{ldA}{256},
        \overparameq{B}{$B$},
        \overparameq{ldB}{512})%
    },
\end{center}
corresponding to the operation $B \leftarrow 0.37 B A^{-1}$, where
$A \in \mathbb R^{128 \times 128}$ is lower triangular with leading
dimension $\mathtt{ldA} = 256$, and $B \in \mathbb R^{512 \times 128}$
with $\mathtt{ldB} = 512$. In our experiment, 
this operation is repeatedly executed on
one core of our Harpertown, using the high-performance BLAS 
implementations \diff{OpenBLAS}, MKL, and ATLAS\footnote{
    \diff{BLAS provides us with the necessary kernels used in the calculations; we do
    not attempt any optimization of such routines.}
}. 
Notice that the first invocation of a BLAS routine is always
notably (in our case more than one order of magnitude) slower than the following one,
due to the initialization of BLAS, which happens at the first invocation of the library.
Neglecting these first measurement outliers, 
the performance measurements of the routine executions with both
in-cache and out-of-cache arguments are shown in
\autoref{fig:sampling.experiments.plot}.
As expected, in-cache corresponds to higher performance across all
implementations, while the increase in execution time for out-of-cache
varies from one implementation to the other. 
In our study ---in which the performance of algorithms is obtained through models of
the algorithms' components--- memory locality will play a big role.

In addition to the influence of memory locality, we observe fluctuations in the
performance measurements of about $8\%$. For this reason, we do not consider
the routine's performance to be one number but a probabilistic distribution. To
express the performance in numbers, we select certain properties of this
distribution, such as minimum, average, standard deviation, and median.

        \subsection{The Sampler}
        \label{sec:sampling.tool}
        To facilitate the acquisition of  performance measurements, 
we wrote the Sampler, a flexible lightweight performance measurement tool. 
Written in C, the Sampler directly
interfaces with libraries such as BLAS or LAPACK. Its configuration allows to
choose between different memory locality situations. Given routine names and
arguments in the form of tuples, such as $(dtrsm, R, L, N , U, 512, 128, 0.37, A,
256, B, 512)$ ($A$ and $B$ specify the sizes of the operands), the Sampler measures
and reports the performance of the routines; this entails collecting multiple samples 
and extracting statistical information.

    \section{Modeling}
    \label{sec:modeling}
    With the measurements obtained by the Sampler, we want now to construct analytical
performance models. Here we introduce the \emph{Modeler}, a tool which interacts with
the Sampler and automatically generates performance models. 
These models form the base for the performance prediction and algorithm ranking 
(\autoref{sec:ranking}).

        \subsection{Preliminary Experiments}
        \label{sec:modeling.experiments}
        Most BLAS routines accept 10 or more arguments; LAPACK's routines have
easily twice as many. In building performance models, if we 
blindly treated all the arguments equally, we would
originate 10+ dimensional models, which would result in either 
impractical execution times or sloppy accuracy.
To avoid this curse of dimensionality, we analyze how different
arguments types affect performance, and in our models we only account
for a subset of the arguments.

Here we focus on the dependence of performance on the BLAS arguments.
Again, we use \texttt{dtrsm} as an example;
the arguments of BLAS routines can be classified as follows:
\begin{center}
    \newcommand{\overparambrace}[2]{%
        \tikz[baseline=(n.base)] {
            \node[inner sep=0pt, label=90:] (n) {\texttt{\vphantom{gT}#2}};
            \draw[decorate, decoration=brace] (n.north west) -- (n.north east) node[midway, above] {\textnormal{\vphantom{Tg}#1}};
        }%
    }
    \small
    \texttt{%
        dtrsm(%
        \overparambrace{flag}{side, uplo, transA, diag},
        \overparambrace{size}{m, n},
        \\
        \tikz[baseline=(A.base)] {
            \node (pos) {};
            \def\oldlabel{pos}
            \foreach \label/\text in {
                    scal/{alpha}, c/{,\ \ },
                    A/{A}, c/{,\ \ },
                    ldA/{ldA}, c/{,\ \ },
                    B/{B}, c/{,\ \ },
                    ldB/{ldB}, tail/{)\textnormal{.}}
                } {
                \node[inner sep=0, anchor=base west, base right=0 of \oldlabel.base east] (\label) {\texttt{\vphantom{p}\text}};
                \xdef\oldlabel{\label}
            }
            \draw[decorate, decoration=brace] (scal.north west) -- (scal.north east) node[midway, above] {\textnormal{\vphantom{gT}scalar}};
            \path (A.north) -- (B.north) node[midway, above] (data) {\textnormal{\vphantom{gT}{data}}};
            \draw[->] (data) -- (A.north east |- 0, .25);
            \draw[->] (data) -- (B.north west |- 0, .25);
            \path (ldA.south) -- (ldB.south) coordinate[midway, below] (ld);
            \path (ld) ++(0, -.1) node[anchor=north] (ld) {\textnormal{\vphantom{gT}{leading dimension}}};
            \draw[->] (ld) -- (ldA.south);
            \draw[->] (ld) -- (ldB.south);
        }%
    }
\end{center}
\emph{Flag} arguments take one of only two values (e.g., $\mathtt{side} \in \{\mathtt L, \mathtt R\}$);
\emph{size} arguments contain the dimensions of the matrix and vector operands;
\emph{scalar}s are floating point numbers, which scale the operands;
\emph{data} are (pointers to) the buffers in which the operands are stored;
\emph{leading dimension}s define the distance in memory between two horizontally adjacent matrix entries.

Due to the following reasons,
for our purposes we can disregard all but flag and size arguments. 
\begin{itemize}
    \item Scalar arguments are usually set to $1$ or $-1$. Neither of these values requires any floating point operations to perform the scaling, thus not affecting performance.
    Even other values for scalar arguments have an insignificant influence on performance, since they only affect a lower order term of the operation count.
    
    \item As discussed in~\autoref{sec:sampling.goal.input}, only the size and
    storage location of vector and matrix arguments are relevant for
    performance. The sizes of these operands are covered by the size and leading
    dimension arguments. As for the storage locations, we will construct
    separate models for different memory locality scenarios, so that we can
    entirely ignore data arguments within one model.
    
    \item In practice, the leading dimensions are either equal to the size
    of a corresponding input matrix or larger. While the
    difference between these two scenarios can influence performance, we only
    need to consider the latter: within our targeted algorithms, the BLAS
    routines are invoked on parts of a large input matrix of constant
    dimension. Hence, throughout the model generation, all leading dimension
    arguments are set to $2500$.
\end{itemize}

Next, we study the influence of the flags and the size arguments 
on the performance of BLAS routines.

            \subsubsection{Flag Arguments}
            \label{sec:modeling.experiments.disc}
            All types of flag arguments encountered 
in BLAS appear in the signature of \texttt{dtrsm}:
$\mathtt{side} \in \{\mathtt L, \mathtt R\}$ defines from which side $B$ is multiplied by $A^{-1}$;
$\mathtt{uplo} \in \{\mathtt L, \mathtt U\}$ states if $A$ is lower or upper triangular;
$\mathtt{transA} \in \{\mathtt N, \mathtt T\}$ indicates whether $A$ or its transpose $A^T$ is to be used;
when set to \texttt U, $\mathtt{diag} \in \{\mathtt N, \mathtt U\}$ declares that $A$ is unit triangular.

In \autoref{fig:modeling.experiments.disc.plot},
we report on a series of experiments in which we look at the performance 
for all possible combinations of the flags; the remaining arguments are fixed as follows:
\begin{center}
    \small
    \texttt{%
        dtrsm(%
        \overparameq{side}{\textit{side}},
        \overparameq{uplo}{\textit{uplo}},
        \overparameq{transA}{\textit{transA}},
        \overparameq{diag}{\textit{diag}},
        \overparameq{m}{256},
        \overparameq{n}{256},
        \overparameq{alpha}{0.5},
        \overparameq{A}{$A$},
        \overparameq{ldA}{256},
        \overparameq{B}{$B$},
        \overparameq{ldB}{256})%
    }.
\end{center}

\diff{\begin{figure}[t]
    \scriptsize
    \centering
    \tikzset{external/export=true}

    \begin{tikzpicture}
        \begin{axis}[
            onecolplot,
            height=.3\textwidth,
            ylabel={\metric{ticks}},
            legend columns=-1,
            legend pos=south west,
            xmin=0.5,
            xmax=16.5,
            xtick={1,2,3,4,5,6,7,8,9,10,11,12,13,14,15,16},
            xticklabels={
                \texttt{L L N N},
                \texttt{L L N U},
                \texttt{L L T N},
                \texttt{L L T U},
                \texttt{L U N N},
                \texttt{L U N U},
                \texttt{L U T N},
                \texttt{L U T U},
                \texttt{R L N N},
                \texttt{R L N U},
                \texttt{R L T N},
                \texttt{R L T U},
                \texttt{R U N N},
                \texttt{R U N U},
                \texttt{R U T N},
                \texttt{R U T U}
            },
            every x tick label/.append style={text width=.3cm, align=center},
            ymin=0
        ]
            \addplot[color=plot1, only marks] file {figures/data/modeling.experiments.disc.plot/open.ticks.dat};
            \label{fig:modeling.experiments.disc.plot:open}
            \addlegendentry{OpenBLAS}

            \addplot[color=plot2, only marks] file {figures/data/modeling.experiments.disc.plot/mkl.ticks.dat};
            \label{fig:modeling.experiments.disc.plot:mkl}
            \addlegendentry{MKL}

            \addplot[color=plot3, only marks] file {figures/data/modeling.experiments.disc.plot/atlas.ticks.dat};
            \label{fig:modeling.experiments.disc.plot:atlas}
            \addlegendentry{ATLAS}
        \end{axis}
        \draw (0, 0) node[anchor=north east, align=left] {
            \texttt{side}\\
            \texttt{uplo}\\
            \texttt{transA}\\
            \texttt{diag}
        };
    \end{tikzpicture}

    \caption{\texttt{dtrsm}: \metric{ticks} as a function of the discrete arguments.}
    \label{fig:modeling.experiments.disc.plot}
    \tikzset{external/export=false}
\end{figure}}

The only common feature across all
implementations is that \texttt{diag} only has a minor impact
on performance. 
No clear pattern arises to relate the performance of two or more arguments. 
This may be due to different argument values leading to the execution of 
distinct code branches.
We conclude that in our models, with the exception of \texttt{diag},
we should treat all combinations of argument values separately.

            \subsubsection{Size Arguments}
            \label{sec:modeling.experiments.size}
            \diff{ \begin{figure}[t]
    \scriptsize
    \centering
    \tikzset{external/export=true}

    \begin{tikzpicture}
        \begin{axis}[
            onecolplot,
            xlabel={$\mathtt n$},
            ylabel={\metric{ticks}},
            xtick={8,256,512,...,1024},
            xmin={},
            ymin={},
            legend pos=north west
        ]
            \addlegendimage{plot1, only marks}
            \addlegendentry{OpenBLAS}
            \label{fig:modeling.experiments.size:open}
            \addlegendimage{plot2, only marks}
            \addlegendentry{MKL}
            \addlegendimage{plot3, only marks}
            \addlegendentry{ATLAS}

            \addplot[color=plot1, mark size=.3pt, only marks] table[x index=5, y index=10, header=false] {figures/data/modeling.experiments.size/open.size.dat};
            \addplot[color=plot2, mark size=.3pt, only marks] table[x index=5, y index=10, header=false] {figures/data/modeling.experiments.size/mkl.size.dat};
            \addplot[color=plot3, mark size=.3pt, only marks] table[x index=5, y index=10, header=false] {figures/data/modeling.experiments.size/atlas.size.dat};

        \end{axis}
    \end{tikzpicture}

    \caption{\texttt{dgemm}: \metric{ticks} as a function of the size arguments.}
    \label{fig:modeling.experiments.size.ticks}
    \tikzset{external/export=false}
\end{figure}

\begin{figure}[t]
    \scriptsize
    \centering
    \tikzset{external/export=true}

    \begin{tikzpicture}
        \begin{axis}[
            onecolplot,
            xlabel={$\mathtt n$},
            ylabel={$\metric{ticks} - p$},
            xtick={8,256,512,...,1024},
            ymin=-3e6,
            ymax=3e6,
            xmin={},
            legend pos=north west 
        ]
            \addlegendimage{plot1, only marks}
            \addlegendentry{OpenBLAS}
            \label{fig:modeling.experiments.size:open}
            \addlegendimage{plot2, only marks}
            \addlegendentry{MKL}
            \addlegendimage{plot3, only marks}
            \addlegendentry{ATLAS}

            \addplot[color=plot1, mark size=.3pt, only marks] table[x index=5, header=false, y expr=\thisrowno{10} - (275532.595736 + -936.906151751*\thisrowno5 + 2.94119654815*\thisrowno5^2 + 0.290605001261*\thisrowno5^3)] {figures/data/modeling.experiments.size/open.size.dat};
            \addplot[color=plot2, mark size=.3pt, only marks] table[x index=5, header=false, y expr=\thisrowno{10} - (-1623072.7436 + 25216.5963562*\thisrowno5 + -72.3929302054*\thisrowno5^2 + 0.331118281956*\thisrowno5^3)] {figures/data/modeling.experiments.size/mkl.size.dat};
            \addplot[color=plot3, mark size=.3pt, only marks] table[x index=5, header=false, y expr=\thisrowno{10} - (-284559.515161 + 6401.38675266*\thisrowno5 + -2.64887587434*\thisrowno5^2 + 0.318300376623*\thisrowno5^3)] {figures/data/modeling.experiments.size/atlas.size.dat};
        \end{axis}
    \end{tikzpicture}

    \caption{\texttt{dgemm}: Distance between least-squares fitting and original data (\autoref{fig:modeling.experiments.size.ticks}).}
    \label{fig:modeling.experiments.size.error}
    \tikzset{external/export=false}
\end{figure}}

We consider the invocation
\begin{center}
    \small
    \texttt{%
        dtrsm(%
        \overparameq{side}{L},
        \overparameq{uplo}{L},
        \overparameq{transA}{N},
        \overparameq{diag}{N},
        \overparameq{m}{n},
        \overparameq{n}{n},
        \\
        \overparameq{alpha}{0.5},
        \overparameq{A}{$A$},
        \overparameq{ldA}{n},
        \overparameq{B}{$B$},
        \overparameq{ldB}{n}),%
    }
\end{center}
where \texttt n varies between 8 and 1024.
To avoid the influence of small
scale fluctuation, we only consider values of \texttt n that are multiples of
8. Measurements for different BLAS libraries 
are shown in \autoref{fig:modeling.experiments.size.ticks}.

At first sight, the measurements follow a quadratic behavior,
in line with the routine's complexity. For each of the three BLAS implementations, 
we construct a quadratic polynomial $p$ that best approximates the measurements
through least squares fitting. 
In \autoref{fig:modeling.experiments.size.error}
we plot the difference between $p$
and the original measurements;
it becomes apparent that in none of the three cases, a quadratic 
approximation represents the performance accurately.
However, the plot shows some degree of structure; this is
especially visible for \diff{OpenBLAS}~(\ref{fig:modeling.experiments.size:open}),
where there are intervals with polynomial behavior separated by jumps or
kinks. We can make out the following intervals: \diff{8 -- 550, 550 -- 750 and
750 -- 1024}. A similar behavior can be
found in the other implementations, although the higher fluctuations
in their measurements make the jumps less visible. 

This experiment teaches us that a single polynomial cannot accurately
represent the performance of a routine, even when all the arguments
are fixed, and only the size varies. As a consequence, the Modeler
will generate models in the form of piecewise multivariate
polynomials.

        \subsection{The Targeted Models}
        \label{sec:modeling.models}
        A performance model represents the performance of a routine for a fixed
implementation, system, and memory locality situation. Given a set of valid
routine arguments, the model provides estimates on the expected performance in
the form of statistical quantities, such as minimum, average, standard
deviation, and median.

Internally, our models operate as follows. In order to avoid the curse of
dimensionality, only a subset of the routine's arguments are selected; these
are the model \emph{parameters}. We distinguish between two types of parameters:
flags, corresponding to flag arguments, and integer parameters,
corresponding to size (and possibly leading dimension) arguments.

In our models, each combination of flags is treated separately.
In a model with $3$ flag parameters, with $2$ possible values each, this would lead
to $2^3 = 8$ separate \emph{submodels}, representing the performance dependence on
the integer parameters.\footnote{%
    Since there are no more than 4 flag arguments in BLAS routines, the number of submodels stays well within manageable bounds.
}

Each submodel is essentially a vector-valued multivariate piecewise polynomial
in the following sense. The integer parameters span a multidimensional space,
which is covered by rectangular regions, in which the behavior is represented
by polynomials. Each polynomial is vector valued with one value for each
statistical quantity.

When the model is used to estimate the performance for given routine arguments,
the following happens:
(1) the model parameters are extracted;
(2) the submodel corresponding to the combination of flags is identified;
(3) the region containing the integer parameter point is found;
(4) the polynomial corresponding to the region is evaluated, yielding the estimates.

        \subsection{The Modeler}
        \label{sec:modeling.tool}
        In the previous section, we have described the structure of our targeted
performance models. We now introduce the Modeler, a tool that generates
these models automatically.

We skip the technicalities that arise from creating a separate model for each
combination of flags and instead focus on the generation of piecewise
polynomials to model the dependence of performance on integer parameters.
The objective of the Modeler is to attain accuracy automatically and with as few measurements as possible.
Moreover, although within BLAS at most three integer parameters are encountered,
the Modeler is designed for arbitrary dimension.

                \paragraph{Polynomial Fitting through Least Squares}
                \label{sec:modeling.tool.pmodeler.lsfit}
                The approximation of a set of sampling results by polynomials through least
squares fitting is a fundamental task of the modeling process.
A set of $n$ coordinate value pairs $(\mathbf x_i, v_i)$ are fitted with a
polynomial $p$ of limited order, such that
$$
    \sum\limits_{i=1}^n (p(\mathbf x_i) - v_i)^2
$$
is minimized. To solve this least squares problem, we use the function
\texttt{linalg.lstsq()} provided by Python's SciPy package, which is based on
singular value decomposition.

The accuracy of a polynomial approximation $p$ is determined by the local
errors $e_i = p(\mathbf x_i) - v_i$. While the used least squares method
minimizes $\sum\limits_{i = 1}^n e_i^2$, we use the \emph{maximum relative
error} across all $\mathbf x_i$:
$$
    e_\text{relmax} = \max\limits_{1 \leq i \leq n} \frac{|e_i|}{v_i}.
$$

To create the vector valued polynomial for the statistical quantities, each
quantity is separately fitted with a polynomial. The error $e_\text{relmax}$ of
one quantity is picked to represent the accuracy of the whole polynomial.
In the following, we use the median for this purpose.

            \subsubsection{Model Expansion}
            \label{sec:modeling.tool.exp}
            We now introduce the first of two modeling strategies, \emph{Model Expansion}. 
The piecewise polynomial is created according to the following steps.
\begin{itemize}
    \item The first objective is to build a model for 
    a small region in a corner of the parameter
    space; this is accomplished by fitting 
    a small set of measurements.
    \item This initial region is then expanded as much as possible, 
    by taking new measurements, integrating them in the model, and checking that
    \begin{itemize}
        \item[a)] the polynomial's approximation error is below a given threshold, 
        and
        \item[b)] the region stays within the boundaries of the parameter space.
    \end{itemize}
    \item Once a region cannot be extended further ---because of either a) or b)--- 
    new adjacent regions are generated and expanded. 
\end{itemize}
The process is repeated until the whole parameter space is covered. 

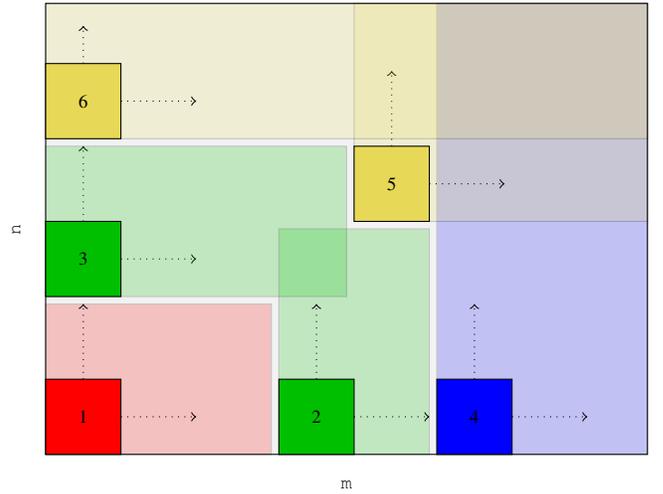
\begin{figure}[t]
    \scriptsize
    \centering
    \tikzset{external/export=true}
    
    \begin{tikzpicture}
        \filldraw[fill=graybg] (0, 0) rectangle (8, 6);
        \node[anchor=north] at (4, -.25) {\texttt m};
        \node[anchor=south, rotate=90] at (-.25, 3) {\texttt n};

        \filldraw[fill=plot1, opacity=.2] (0, 0) rectangle ++(3, 2);
        \filldraw[fill=plot1] (0, 0) rectangle ++(1, 1) node[midway] {1};
        \draw[->, dotted] (.5, 1) -- ++(0, 1);
        \draw[->, dotted] (1, .5) -- ++(1, 0);
        
        \filldraw[fill=plot2, opacity=.2] (3.1, 0) rectangle ++(2, 3);
        \filldraw[fill=plot2] (3.1, 0) rectangle ++(1, 1) node[midway] {2};
        \draw[->, dotted] (.5, 3.1) -- ++(0, 1);
        \draw[->, dotted] (1, 2.6) -- ++(1, 0);
        \filldraw[fill=plot2, opacity=.2] (0, 2.1) rectangle ++(4, 2);
        \filldraw[fill=plot2] (0, 2.1) rectangle ++(1, 1) node[midway] {3};
        \draw[->, dotted] (3.6, 1) -- ++(0, 1);
        \draw[->, dotted] (4.1, .5) -- ++(1, 0);

        \filldraw[fill=plot3, opacity=.2] (5.2, 0) rectangle (8, 6);
        \filldraw[fill=plot3] (5.2, 0) rectangle ++(1, 1) node[midway] {4};
        \draw[->, dotted] (5.7, 1) -- ++(0, 1);
        \draw[->, dotted] (6.2, .5) -- ++(1, 0);

        \filldraw[fill=plot4, opacity=.2] (4.1, 3.1) rectangle (8, 6);
        \filldraw[fill=plot4] (4.1, 3.1) rectangle ++(1, 1) node[midway] {5};
        \draw[->, dotted] (4.6, 4.1) -- ++(0, 1);
        \draw[->, dotted] (5.1, 3.6) -- ++(1, 0);
        \filldraw[fill=plot4, opacity=.2] (0, 4.2) rectangle (8, 6);
        \filldraw[fill=plot4] (0, 4.2) rectangle ++(1, 1) node[midway] {6};
        \draw[->, dotted] (.5, 5.2) -- ++(0, .5);
        \draw[->, dotted] (1, 4.7) -- ++(1, 0);
    \end{tikzpicture}
    
    \tikzset{external/export=false}
    \caption{Sequence of steps in the construction of piecewise models through Model Expansion.}
    \label{fig:modeling.tool.exp}
\end{figure}

An example of Model Expansion in two dimensions is given in
\autoref{fig:modeling.tool.exp}. The rectangular domain is filled 
starting with region 1 (\tikzsquare{plot1}) in the bottom-left corner 
of the domain. This region is expanded in both directions until its 
accuracy reaches the threshold (\tikzsquare{plot1, opacity=.2}). 
Two new adjacent regions 2 and 3 (\tikzsquare{plot2}) are then created 
---together with the associated samples--- and their expansion begins. 
Assuming that region 2 was expanded as far as possible 
(\tikzsquare{plot2, opacity=.2}), region 4 (\tikzsquare{plot3}) is then 
generated. Once the expansion for region 3 (\tikzsquare{plot2}) is complete, 
the neighboring regions 5 and 6 (\tikzsquare{plot4}) are created. These 6 models 
cover the full parameter space, therefore the process terminates.

            \subsubsection{Adaptive Refinement}
            \label{sec:modeling.tool.ref}
            The second strategy to generate piecewise models is based on adaptive refinement. 
The idea is to begin with a simple and regular model constructed from a coarse 
grid of samples across the whole parameter space; the quality of such a model is 
then evaluated. If insufficient, the region is split and the model is refined 
by locally increasing the sample grid resolution. These steps are applied 
recursively to the refined regions until either the accuracy reaches a satisfactory 
level across the whole domain, or a given resolution limit is reached.

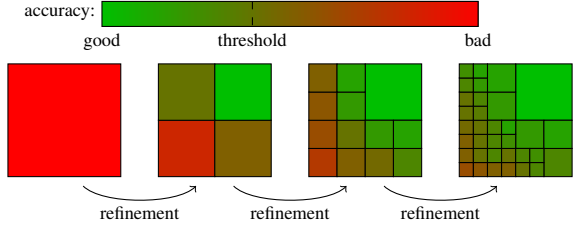
\begin{figure}[t]
    \scriptsize
    \centering
    
    \begin{tikzpicture}[yscale=.666]
         \coordinate (pos) at (0, 0);
         \filldraw[fill=plot2] (pos) +(-2.5, 0) rectangle +(2.5, -.5);
         \filldraw[path fading=west, fill=plot1] (pos) +(-2.5, 0) rectangle +(2.5, -.5);
         \path (pos) ++(-2.5, -.25) node[anchor=east] {accuracy:};
         \path (pos) ++(-2.5, -.5) node[anchor=north] {good};
         \path (pos) ++(2.5, -.5) node[anchor=north] {bad};
         \draw[dashed] (pos) ++(-2.5, 0) ++($.4*(5, 0)$) -- ++(0, -.5) node[anchor=north] {threshold};

         \path (pos) ++(2.5, -.25) node[anchor=west] {\phantom{accuracy:}};
    \end{tikzpicture}

    \vspace{.1cm}

    \tikzset{external/export=true}
    \begin{tikzpicture}[scale=.5]
        \coordinate (pos) at (0, 0);
        \filldraw[fill=plot1!100!plot2] (pos) rectangle ++(3, 3);

        \path (pos) ++(4, 0) coordinate (pos);

        \foreach \x/\y/\c in {
            0/0/80,     .5/0/50,
            0/.5/40,    .5/.5/0}
            \filldraw[fill=plot1!\c!plot2] (pos) ++($3*(\x, \y)$) rectangle ++(1.5, 1.5);

        \path (pos) ++(4, 0) coordinate (pos);

        \foreach \x/\y/\c in {
                        .5/.5/0}
            \filldraw[fill=plot1!\c!plot2] (pos) ++($3*(\x, \y)$) rectangle ++(1.5, 1.5);
        \foreach \x/\y/\c in {
            0/0/70,     .25/0/50,   .5/0/45,    .75/0/30,
            0/.25/60,   .25/.25/40, .5/.25/20,  .75/.25/15,
            0/.5/55,    .25/.5/20,
            0/.75/50,   .25/.75/15}
            \filldraw[fill=plot1!\c!plot2] (pos) ++($3*(\x, \y)$) rectangle ++(.75, .75);

        \path (pos) ++(4, 0) coordinate (pos);

        \foreach \x/\y/\c in {
                        .5/.5/0}
            \filldraw[fill=plot1!\c!plot2] (pos) ++($3*(\x, \y)$) rectangle ++(1.5, 1.5);
        \foreach \x/\y/\c in {
                                                .75/0/30,
                                    .5/.25/20,  .75/.25/15,
                        .25/.5/20,
                        .25/.75/15}
            \filldraw[fill=plot1!\c!plot2] (pos) ++($3*(\x, \y)$) rectangle ++(.75, .75);
        \foreach \x/\y/\c in {
            0/0/60,     .125/0/55,      .25/0/50,       .375/0/50,      .5/0/40,    .625/0/30,
            0/.125/50,  .125/.125/40,   .25/.125/30,    .375/.125/40,   .5/.125/30, .625/.125/25,
            0/.25/50,   .125/.25/40,    .25/.25/30,     .375/.25/25,
            0/.375/45,  .125/.375/40,   .25/.375/30,    .375/.375/15,
            0/.5/40,    .125/.5/35,
            0/.625/40,  .125/.625/30,
            0/.75/30,   .125/.75/20,
            0/.875/25,  .125/.875/15}
            \filldraw[fill=plot1!\c!plot2] (pos) ++($3*(\x, \y)$) rectangle ++(.375, .375);

         \coordinate (pos) at (2, -.25);
         \draw[->] (pos) .. controls ++(.5, -.5) and ++(-.5, -.5) .. ++(3, 0) node[midway, below] {refinement};
         \path (pos) ++(4, 0) coordinate (pos);
         \draw[->] (pos) .. controls ++(.5, -.5) and ++(-.5, -.5) .. ++(3, 0) node[midway, below] {refinement};
         \path (pos) ++(4, 0) coordinate (pos);
         \draw[->] (pos) .. controls ++(.5, -.5) and ++(-.5, -.5) .. ++(3, 0) node[midway, below] {refinement};
    \end{tikzpicture}

    \tikzset{external/export=false}
    \caption{Sequence of steps in the construction of piecewise models through Adaptive Refinement.}
    \label{fig:modeling.tool.ref.refine}
\end{figure}

An example of Adaptive Refinement for a two dimensional domain is shown in
\autoref{fig:modeling.tool.ref.refine}. The polynomial approximation for the
initial region spanning the entire parameter space is very inaccurate
(\tikzsquare{plot1}, 1\textsuperscript{st} square on the left). Therefore it is refined,
generating four new regions, and new measurements are obtained to 
create four polynomials (2\textsuperscript{nd} square on the left). Now,
the error in the top right quadrant (\tikzsquare{plot2}) is already below the
threshold (\tikzsquare{plot1!40!plot2}); the other quadrants are not accurate
enough and are further refined (3\textsuperscript{rd} square). In the next
iteration, several regions are below the desired error threshold; the others 
are refined once more (4\textsuperscript{th} square). Although some of the resulting
polynomials are still above the desired level of accuracy, they are accepted 
anyway, because their size does not allow further refinement.

        \subsection{Results}
        \label{sec:modeling.res}
        Having introduced two modeling strategies, 
here we compare the resulting models, both in terms of speed and accuracy.
We consider again the solution of a triangular system
$B \leftarrow A^{-1} B$ as testbed:
\begin{center}
    \small
    \texttt{dtrsm(side, uplo, transA, diag, side, m, n, alpha, A, ldA, B, ldB)}.
\end{center}
The interface of this routine contains four flags (\texttt{side} through
\texttt{diag}), two size arguments (\texttt m and \texttt n), one scalar
argument (\texttt{alpha}), and operates on two matrices (\texttt A and \texttt
B with corresponding leading dimensions \texttt{ldA} and \texttt{ldB}). 
Out of these arguments, our models account for
\begin{itemize}
    \item the flag parameters \texttt{side}, \texttt{uplo}, and \texttt{transA}, and
    \item the integer parameters \texttt m and \texttt n.
\end{itemize}

The integer parameters vary in the range $[8 - 1024]$ and 
define the parameter space; 
the flags are $(\mathtt{side}, \mathtt{uplo},
\mathtt{transA}) = (\mathtt L, \mathtt L, \mathtt N)$;
the values of the remaining arguments are: $\mathtt{diag} =
\mathtt N$, $\mathtt{alpha} = 0.5$, and $\mathtt{ldA} = \mathtt{ldB} = 2500$.
We use the in-cache configuration of the Sampler and \diff{OpenBLAS} \diff{on the Harpertown processor}.

            \subsubsection{Model Expansion}
            \label{sec:modeling.res.me}
            This approach  accepts several configuration options:
\begin{itemize}
    \item the relative error bound $\varepsilon$;

    \item the direction of expansion $d \in \{\nearrow, \swarrow\}$;

    \item the initial size of regions $s_{ini}$.
\end{itemize}

\diff{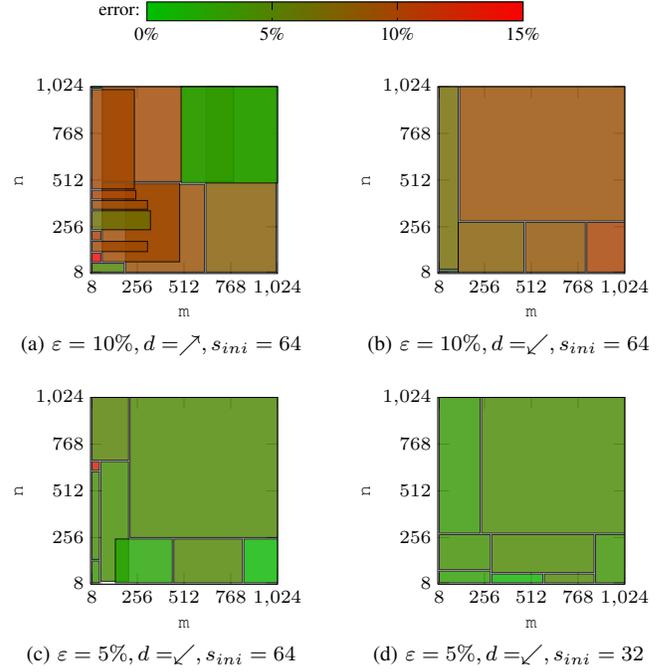
\begin{figure}[t]
    \scriptsize
    \centering
    \begin{tikzpicture}[yscale=.5]
        \coordinate (pos) at (0, 0);
        \filldraw[fill=plot2] (pos) +(-2.5, 0) rectangle +(2.5, -.5);
        \filldraw[path fading=west, fill=plot1] (pos) +(-2.5, 0) rectangle +(2.5, -.5);
        \path (pos) ++(-2.5, -.25) node[anchor=east] {error:};
        \path (pos) ++(2.5, -.25) node[anchor=west] {\phantom{error:}};
        \path (pos) ++(-2.5, -.5) node[anchor=north] {0\%};
        \path (pos) ++(2.5, -.5) node[anchor=north] {15\%};
        \draw[dashed] (pos) ++(-2.5, 0) ++($.666*(5, 0)$) -- ++(0, -.5) node[anchor=north] {10\%};
        \draw[dashed] (pos) ++(-2.5, 0) ++($.333*(5, 0)$) -- ++(0, -.5) node[anchor=north] {5\%};
    \end{tikzpicture}

    \tikzset{external/export=true}
    \newcommand{\regionsplot}[1]{
        \begin{tikzpicture}
            \begin{axis}[
                twocolplot,
                axis equal image=true,
                xlabel={\texttt m},
                ylabel={\texttt n},
                xtick={8,256,512,...,1024},
                ytick={8,256,512,...,1024},
                xmax=1024,
                ymax=1024
            ]
                \coordinate (oi) at (axis description cs: 0, 1);
                \coordinate (io) at (axis description cs: 1, 0);
            \end{axis}

            \foreach \ax/\ay/\bx/\by/\e in {#1} {
                \pgfmathparse{100 * max(0, min(1, 1 * (1 - \e / .15)))}
                \filldraw[fill=plot2!\pgfmathresult!plot1, opacity=.8] ($\ax/1024*(io) + \ay/1024*(oi)$) rectangle ($\bx/1024*(io) + \by/1024*(oi)$);
            }
        \end{tikzpicture}
    }
    \subfloat[$\varepsilon= 10\%, d= \nearrow, s_{ini} = 64$]{
        \label{fig:modeling.res.ticks.me.regions:a10.sg32.dr}
        \regionsplot{
            8/8/184/56/0.0461577492863,
            192/8/624/488/0.0896726476306,
            8/64/56/112/0.167461245102,
            64/64/488/496/0.093744629853,
            8/120/312/176/0.0938512491326,
            8/184/56/232/0.102666263604,
            8/240/328/344/0.063217932469,
            632/8/1016/1024/0.0775137380107,
            64/504/784/1024/0.0945803711394,
            496/496/1024/1024/0.0136969095611,
            8/352/312/400/0.0915324548666,
            8/408/248/456/0.0993595388281,
            8/464/240/1008/0.0939803539301,
            8/1016/56/1024/0.0014294926915
        }
    }
    \hfill
    \subfloat[$\varepsilon= 10\%, d= \swarrow, s_{ini} = 64$]{
        \label{fig:modeling.res.ticks.me.regions:a10.sg32.dl}
        \regionsplot{
            120/288/1024/1024/0.0888675571342,
            8/24/112/1024/0.0668288392833,
            816/8/1024/280/0.0995070701817,
            480/8/808/280/0.0854691082839,
            8/8/112/16/0.0194043043893,
            112/8/472/280/0.0772282959808
        }
    }

    \subfloat[$\varepsilon= 5\%, d= \swarrow, s_{ini} = 64$]{
        \label{fig:modeling.res.ticks.me.regions:a5.sg32.dl}
        \regionsplot{
            216/256/1024/1024/0.0441692906432,
            840/8/1024/248/0.0052610533895,
            8/680/208/1024/0.0497228902249,
            456/8/832/248/0.0458701584706,
            56/16/208/672/0.0391441374965,
            136/8/448/248/0.0205056341004,
            8/624/48/672/0.12603031663,
            8/136/48/616/0.0393446730635,
            8/8/48/128/0.0323501577701
        }
    }
    \hfill
    \subfloat[$\varepsilon= 5\%, d= \swarrow, s_{ini} = 32$]{
        \label{fig:modeling.res.ticks.me.regions:a5.sg16.dl}
        \regionsplot{
            240/280/1024/1024/0.04252291301,
            8/280/232/1024/0.031445145391,
            864/8/1024/272/0.0355253794589,
            296/64/856/272/0.0441857380411,
            8/80/288/272/0.0367278771516,
            584/8/856/56/0.044463515776,
            8/8/288/72/0.0262729258477,
            296/8/576/56/0.0082225274684
        }
    }
    \tikzset{external/export=false}

    \caption{Model Expansion for \texttt{dtrsm}.}
    \label{fig:modeling.res.ticks.me.regions}
\end{figure}
}

We illustrate the influence of such options on the generation by presenting models 
obtained with different settings. The plots in
\autoref{fig:modeling.res.ticks.me.regions} show how differently 
these models cover the parameter space and display the relative errors.

In \autoref{fig:modeling.res.ticks.me.regions:a10.sg32.dr}, we used the
configuration:
\begin{itemize}
    \item the error bound is $\varepsilon = 10\%$;

    \item the direction of expansion is $d = \nearrow$;

    \item new regions are initially of size $s_{ini} = \diff{64}$.
\end{itemize}
Smaller and less accurately modeled regions are generated towards the left side
of the parameter space. Towards the top right corner, the regions become larger
and the relative error decreases. In this part of the parameter space, we also
find several areas which are modeled by two or more overlapping
regions\footnote{%
    When the model is evaluated at a point covered by multiple regions, the
    most accurate model is selected.
}.

In \autoref{fig:modeling.res.ticks.me.regions:a10.sg32.dl} instead, 
we let the Modeler expand along the direction $d =\, \swarrow$. 
We observe the following changes:
\begin{itemize}
    \item especially towards the top right corner, 
      the generated regions are larger;
    \item these regions are of higher accuracy compared to the previous model, 
      although the error bound was not modified;
    \item fewer regions overlap.
\end{itemize}
The average relative error improves from \diff{$8.26\%$} to \diff{$7.77\%$}, while the number
of required sampling points decreases from \diff{$5280$} to \diff{$2070$}. We noticed that
in general, it is preferable to expand the models towards the origin ($\swarrow$).

In \autoref{fig:modeling.res.ticks.me.regions:a5.sg32.dl}, we reduced the
error bound to $\varepsilon = 5\%$. As a result, the average model error
improves from \diff{$7.77\%$} to \diff{$3.87\%$}. This comes at the cost of an increase in the
number of samples: from \diff{$2070$} to \diff{$2990$}. As in the previous cases, the
accuracy of the models decreases as the parameter values become smaller; the
least accurate models appear for small values of \texttt m.

Finally, in \autoref{fig:modeling.res.ticks.me.regions:a5.sg16.dl} we decreased the size of the initial models from $s = \diff{64}$ to $s = \diff{32}$. 
Interestingly, even though the model now makes use of \diff{only $2710$} samples, the average error \diff{decreases} from \diff{$3.87\%$} to \diff{$3.80\%$}.
\diff{This is due to the fact that the sample points are shifted, and irregularities present in the previous models are not captured.}
From this we conclude that an index of accuracy  generated from  used samples is not always sufficient to capture the global quality of a model.

            \subsubsection{Adaptive Refinement}
            \label{sec:modeling.res.ar}
            \diff{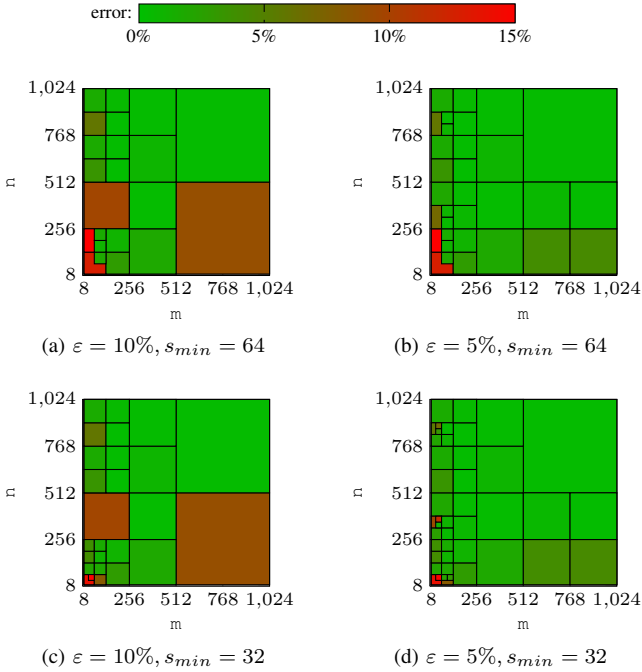
\begin{figure}[t]
    \scriptsize
    \centering
    \begin{tikzpicture}[yscale=.5]
        \coordinate (pos) at (0, 0);
        \filldraw[fill=plot2] (pos) +(-2.5, 0) rectangle +(2.5, -.5);
        \filldraw[path fading=west, fill=plot1] (pos) +(-2.5, 0) rectangle +(2.5, -.5);
        \path (pos) ++(-2.5, -.25) node[anchor=east] {error:};
        \path (pos) ++(2.5, -.25) node[anchor=west] {\phantom{error:}};
        \path (pos) ++(-2.5, -.5) node[anchor=north] {0\%};
        \path (pos) ++(2.5, -.5) node[anchor=north] {15\%};
        \draw[dashed] (pos) ++(-2.5, 0) ++($.666*(5, 0)$) -- ++(0, -.5) node[anchor=north] {10\%};
        \draw[dashed] (pos) ++(-2.5, 0) ++($.333*(5, 0)$) -- ++(0, -.5) node[anchor=north] {5\%};
    \end{tikzpicture}

    \tikzset{external/export=true}
    \newcommand{\regionsplot}[1]{
        \begin{tikzpicture}
            \begin{axis}[
                twocolplot,
                axis equal image=true,
                xlabel={\texttt m},
                ylabel={\texttt n},
                xtick={8,256,512,...,1024},
                ytick={8,256,512,...,1024},
                xmax=1024,
                ymax=1024
            ]
                \coordinate (oi) at (axis description cs: 0, 1);
                \coordinate (io) at (axis description cs: 1, 0);
            \end{axis}

            \foreach \ax/\ay/\bx/\by/\e in {#1} {
                \pgfmathparse{100 * max(0, min(1, 1 * (1 - \e / .15)))}
                \filldraw[fill=plot2!\pgfmathresult!plot1] ($\ax/1024*(io) + \ay/1024*(oi)$) rectangle ($\bx/1024*(io) + \by/1024*(oi)$);
            }
        \end{tikzpicture}
    }
    \subfloat[$\varepsilon = 10\%, s_{min} = 64$]{
        \label{fig:modeling.res.ticks.ar.regions:a10.sg16}
        \regionsplot{
            8/8/1024/1024/22.8998289596,
            8/8/512/512/6.31349340074,
            8/512/512/1024/3.06961796737,
            512/8/1024/512/0.0880091110556,
            512/512/1024/1024/0.00317333958322,
            8/8/256/256/0.521487639129,
            8/256/256/512/0.0954104070152,
            256/8/512/256/0.0177113122687,
            256/256/512/512/0.00291129486899,
            8/512/256/768/0.108561484493,
            8/768/256/1024/0.161461075741,
            256/512/512/768/0.00775959559329,
            256/768/512/1024/0.00298316639008,
            8/8/128/128/0.126454363107,
            8/128/128/256/0.174024389961,
            128/8/256/128/0.0273903213813,
            128/128/256/256/0.00917310480119,
            8/512/128/640/0.0328419239588,
            8/640/128/768/0.0157228089733,
            128/512/256/640/0.00162643543776,
            128/640/256/768/0.00239248283572,
            8/768/128/896/0.0574310577781,
            8/896/128/1024/0.0152696010591,
            128/768/256/896/0.00161539242936,
            128/896/256/1024/0.00394697755889,
            64/64/128/128/0.0176533660849,
            64/128/128/192/0.00895728803614,
            64/192/128/256/0.0147144541737
        }
    }
    \hfill
    \subfloat[$\varepsilon = 5\%, s_{min} = 64$]{
        \label{fig:modeling.res.ticks.ar.regions:a5.sg16}
        \regionsplot{
            8/8/1024/1024/22.8998289596,
            8/8/512/512/6.31349340074,
            8/512/512/1024/3.06961796737,
            512/8/1024/512/0.0880091110556,
            512/512/1024/1024/0.00317333958322,
            8/8/256/256/0.521487639129,
            8/256/256/512/0.0954104070152,
            256/8/512/256/0.0177113122687,
            256/256/512/512/0.00291129486899,
            8/512/256/768/0.108561484493,
            8/768/256/1024/0.161461075741,
            256/512/512/768/0.00775959559329,
            256/768/512/1024/0.00298316639008,
            512/8/768/256/0.0396917885149,
            512/256/768/512/0.00551260124991,
            768/8/1024/256/0.0420231346445,
            768/256/1024/512/0.00242388928018,
            8/8/128/128/0.126454363107,
            8/128/128/256/0.174024389961,
            128/8/256/128/0.0273903213813,
            128/128/256/256/0.00917310480119,
            8/256/128/384/0.0675640492753,
            8/384/128/512/0.0184382265091,
            128/256/256/384/0.00604806598444,
            128/384/256/512/0.00270060863923,
            8/512/128/640/0.0328419239588,
            8/640/128/768/0.0157228089733,
            128/512/256/640/0.00162643543776,
            128/640/256/768/0.00239248283572,
            8/768/128/896/0.0574310577781,
            8/896/128/1024/0.0152696010591,
            128/768/256/896/0.00161539242936,
            128/896/256/1024/0.00394697755889,
            64/64/128/128/0.0176533660849,
            64/128/128/192/0.00895728803614,
            64/192/128/256/0.0147144541737,
            64/256/128/320/0.00421875771972,
            64/320/128/384/0.00797487756563,
            64/768/128/832/0.00818887397152,
            64/832/128/896/0.00289434142968
        }
    }

    \subfloat[$\varepsilon = 10\%, s_{min} = 32$]{
        \label{fig:modeling.res.ticks.ar.regions:a10.sg8}
        \regionsplot{
            8/8/1024/1024/22.8998289596,
            8/8/512/512/6.31349340074,
            8/512/512/1024/3.06961796737,
            512/8/1024/512/0.0880091110556,
            512/512/1024/1024/0.00317333958322,
            8/8/256/256/0.521487639129,
            8/256/256/512/0.0954104070152,
            256/8/512/256/0.0177113122687,
            256/256/512/512/0.00291129486899,
            8/512/256/768/0.108561484493,
            8/768/256/1024/0.161461075741,
            256/512/512/768/0.00775959559329,
            256/768/512/1024/0.00298316639008,
            8/8/128/128/0.126454363107,
            8/128/128/256/0.174024389961,
            128/8/256/128/0.0273903213813,
            128/128/256/256/0.00917310480119,
            8/512/128/640/0.0328419239588,
            8/640/128/768/0.0157228089733,
            128/512/256/640/0.00162643543776,
            128/640/256/768/0.00239248283572,
            8/768/128/896/0.0574310577781,
            8/896/128/1024/0.0152696010591,
            128/768/256/896/0.00161539242936,
            128/896/256/1024/0.00394697755889,
            8/8/64/64/0.137699592324,
            8/64/64/128/0.0206911549863,
            64/8/128/64/0.0795594181091,
            64/64/128/128/0.0176533660849,
            8/128/64/192/0.0396387095409,
            8/192/64/256/0.0371132975374,
            64/128/128/192/0.00895728803614,
            64/192/128/256/0.0147144541737,
            32/32/64/64/0.395496288794
        }
    }
    \hfill
    \subfloat[$\varepsilon= 5\%, s_{min} = 32$]{
        \label{fig:modeling.res.ticks.ar.regions:a5.sg8}
        \regionsplot{
            8/8/1024/1024/22.8998289596,
            8/8/512/512/6.31349340074,
            8/512/512/1024/3.06961796737,
            512/8/1024/512/0.0880091110556,
            512/512/1024/1024/0.00317333958322,
            8/8/256/256/0.521487639129,
            8/256/256/512/0.0954104070152,
            256/8/512/256/0.0177113122687,
            256/256/512/512/0.00291129486899,
            8/512/256/768/0.108561484493,
            8/768/256/1024/0.161461075741,
            256/512/512/768/0.00775959559329,
            256/768/512/1024/0.00298316639008,
            512/8/768/256/0.0396917885149,
            512/256/768/512/0.00551260124991,
            768/8/1024/256/0.0420231346445,
            768/256/1024/512/0.00242388928018,
            8/8/128/128/0.126454363107,
            8/128/128/256/0.174024389961,
            128/8/256/128/0.0273903213813,
            128/128/256/256/0.00917310480119,
            8/256/128/384/0.0675640492753,
            8/384/128/512/0.0184382265091,
            128/256/256/384/0.00604806598444,
            128/384/256/512/0.00270060863923,
            8/512/128/640/0.0328419239588,
            8/640/128/768/0.0157228089733,
            128/512/256/640/0.00162643543776,
            128/640/256/768/0.00239248283572,
            8/768/128/896/0.0574310577781,
            8/896/128/1024/0.0152696010591,
            128/768/256/896/0.00161539242936,
            128/896/256/1024/0.00394697755889,
            8/8/64/64/0.137699592324,
            8/64/64/128/0.0206911549863,
            64/8/128/64/0.0795594181091,
            64/64/128/128/0.0176533660849,
            8/128/64/192/0.0396387095409,
            8/192/64/256/0.0371132975374,
            64/128/128/192/0.00895728803614,
            64/192/128/256/0.0147144541737,
            8/256/64/320/0.0247856932202,
            8/320/64/384/0.0920598341187,
            64/256/128/320/0.00421875771972,
            64/320/128/384/0.00797487756563,
            8/768/64/832/0.0197153675938,
            8/832/64/896/0.0538046488421,
            64/768/128/832/0.00818887397152,
            64/832/128/896/0.00289434142968,
            32/32/64/64/0.395496288794,
            64/32/96/64/0.0227804364739,
            96/32/128/64/0.0302661968417,
            32/320/64/352/0.00850712877638,
            32/352/64/384/0.114548351714,
            32/832/64/864/0.0311825406522,
            32/864/64/896/0.0664457438504
        }
    }
    \tikzset{external/export=false}

    \caption{Adaptive Refinement for \texttt{dtrsm}.}
    \label{fig:modeling.res.ticks.ar.regions}
\end{figure}
}

The generation of models is governed
by two options:
\begin{itemize}
    \item the relative error bound $\varepsilon$, and

    \item the minimum region size $s_{min}$.
\end{itemize}
The regions resulting from different values of these options are
shown in \autoref{fig:modeling.res.ticks.ar.regions}.

The first model in \autoref{fig:modeling.res.ticks.ar.regions:a10.sg16} was
generated with an error bound of $\varepsilon = 10\%$ and a minimum region size
of $s = 64$. The result shows an overall distribution of regions similar to
the model in \autoref{fig:modeling.res.ticks.me.regions:a5.sg16.dl} of Model Expansion: 
Smaller and less accurately modeled regions are predominant
for smaller parameter values --- especially for \texttt{m}.
Regions on the finest level are not generated on the lower edges of the parameter space (beginning at 8), since they would be smaller than $s_{min}$.
The seemingly rectangular regions are parts of larger regions that were only partially refined.

In \autoref{fig:modeling.res.ticks.ar.regions:a5.sg16},
the error bound was decreased to $\varepsilon = 5\%$. 
For such an accuracy to be attained, several of the regions from 
the previous model are further refined, especially on the left side. 
The increased number of regions is covered by
\diff{$3560$} samples --- \diff{$970$} more than previously. The higher accuracy
requirement leads to a decrease in the average error from \diff{$7.17\%$} to \diff{$2.41\%$}.

In the next two experiments,
Figures~\ref{fig:modeling.res.ticks.ar.regions:a10.sg8} and
\ref{fig:modeling.res.ticks.ar.regions:a5.sg8},
we decreased the minimum region size to $s = 32$, 
maintaining the error bound $\varepsilon$ to $10\%$ and $5\%$, respectively. 
This leads to the generation
of many tiny regions. The error bound of $10\%$ ($5\%$) leads to an
average error of \diff{$6.20\%$} (\diff{$1.45\%$}) at the cost of \diff{$3070$} (\diff{$4980$})
samples.

            \subsubsection{Comparison}
            \label{sec:modeling.res.comp}
            \diff{\begin{figure}[t]
    \scriptsize
    \centering
    \tikzset{external/export=true}

    \begin{tikzpicture}
        \begin{axis}[
            onecolplot,
            height=.3\textwidth,
            xlabel={\#samples},
            xlabel near ticks,
            ylabel={average error [\%]},
            ylabel near ticks,
            legend columns=1,
            legend style={
                at={(.99, .5)},
                anchor=east
            }
        ]
            \addplot[nodes near coords, color=plot1, only marks, mark=*, point meta=explicit symbolic] coordinates {
                (5280, 0.0826471561727) [(a)] 
                (2070, 0.0777104262729) [(b)] 
                (2990, 0.0386853160576) [(c)] 
                (2710, 0.0379379624625) [(d)] 
            };
            \addlegendentry{Model Expansion}

            \addplot[nodes near coords, color=plot2, only marks, mark=*, point meta=explicit symbolic] coordinates {
                (2590, 0.0716657563351) [(a)] 
                (3070, 0.062028940686) [(b)] 
                (3560, 0.0241093108228) [(c)] 
                (4980, 0.0144724951738) [(d)] 
            };
            \addlegendentry{Adaptive Refinement}

            \addplot[color=plot1, only marks, mark=*] coordinates {
                (8540, 0.0177467843619) 
                (6180, 0.018528641334) 
                (4530, 0.023420853064) 
                (3540, 0.0256511892805) 
                (4320, 0.0315892991069) 
                (2980, 0.0274865310221) 
                (2710, 0.0379379624625) 
                (2990, 0.0386853160576) 
                (2950, 0.0512691861403) 
                (3050, 0.0371153025166) 
                (2680, 0.0535389697086) 
                (2810, 0.0674704531799) 
                (2130, 0.0731296722757) 
                (2320, 0.0731561418693) 
                (2360, 0.0747479104495) 
                (2440, 0.0674006983327) 
                (2070, 0.0777104262729) 
            };

            \addplot[color=plot2, only marks, mark=*] coordinates {
                (7450, 0.008812785629) 
                (4840, 0.0194852055793) 
                (6790, 0.0133115630889) 
                (4660, 0.0235048952189) 
                (5680, 0.0138636735129) 
                (4130, 0.0238063684096) 
                (4980, 0.0144724951738) 
                (3560, 0.0241093108228) 
                (4410, 0.0374509596114) 
                (3360, 0.0470877752605) 
                (3870, 0.062028940686) 
                (3180, 0.0716657563351) 
                (3660, 0.062028940686) 
                (3180, 0.0716657563351) 
                (3360, 0.062028940686) 
                (2880, 0.0716657563351) 
                (3070, 0.062028940686) 
                (2590, 0.0716657563351) 
            };
        \end{axis}
    \end{tikzpicture}

    \caption{Model Expansion vs. Adaptive Refinement.}
    \label{fig:modeling.res.ticks.comp.smpacc}
    \tikzset{external/export=false}
\end{figure}
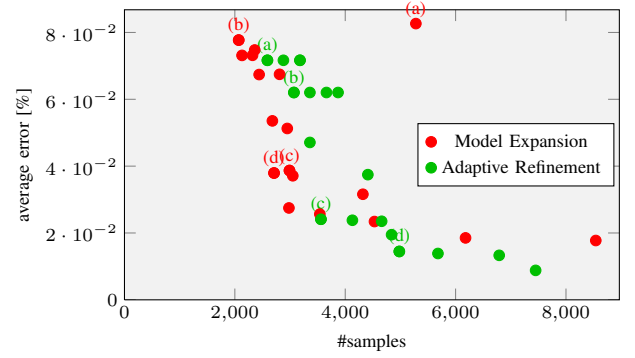}

\autoref{fig:modeling.res.ticks.comp.smpacc} displays, 
for both modeling strategies,
how many samples are needed to generate models of a certain accuracy.
The models presented in the previous sections are labeled according to
Figures~\ref{fig:modeling.res.ticks.me.regions} and
\ref{fig:modeling.res.ticks.ar.regions}.
We are interested in models that attain a high degree of accuracy with
a small number of samples; these are the points laying in the bottom left border
of the convex hull (envelope) of all the points.

For relatively few samples, Model Expansion generates more accurate models
(\textcolor{plot1}{(b)} and \textcolor{plot1}{(d)}). However, one should keep
in mind that these models do not \diff{necessarily} represent the fine scale behavior of
\metric{ticks} very well. If one is willing to use a larger number of samples, Adaptive
Refinement generates more accurate models (\textcolor{plot2}{(c)}). If the number 
of samples is not an issue, this method has the potential to generate very accurate models
(\textcolor{plot2}{(d)}).

In the experiments performed in the rest of the paper,  we used
Adaptive Refinement with configuration \textcolor{plot2}{(c)}: $\varepsilon =
10\%$ error bound and $s_{min} = 32$ minimum regions size. This configuration
is a good compromise between the model accuracy and the number of samples.

    \section{Prediction, Ranking and Optimization}
    \label{sec:ranking}
    We are finally ready to tackle our main goal: ranking linear algebra algorithms by 
performance models. In order to predict the performance of an algorithm, 
we start by analyzing its sequence of subroutine invocations. 
We use the models automatically generated by the
Modeler to estimate the performance of such invocations. These estimates are then
accumulated, resulting in the prediction of the algorithm's performance. The
probabilistic nature of the performance model allows us to give detailed
information on the expected ranges of the algorithm's performance.

        \subsection{Triangular Inverse \texorpdfstring{$L \leftarrow L^{-1}$}{L <- inv(L)}}
        \label{sec:ranking.trinv}
        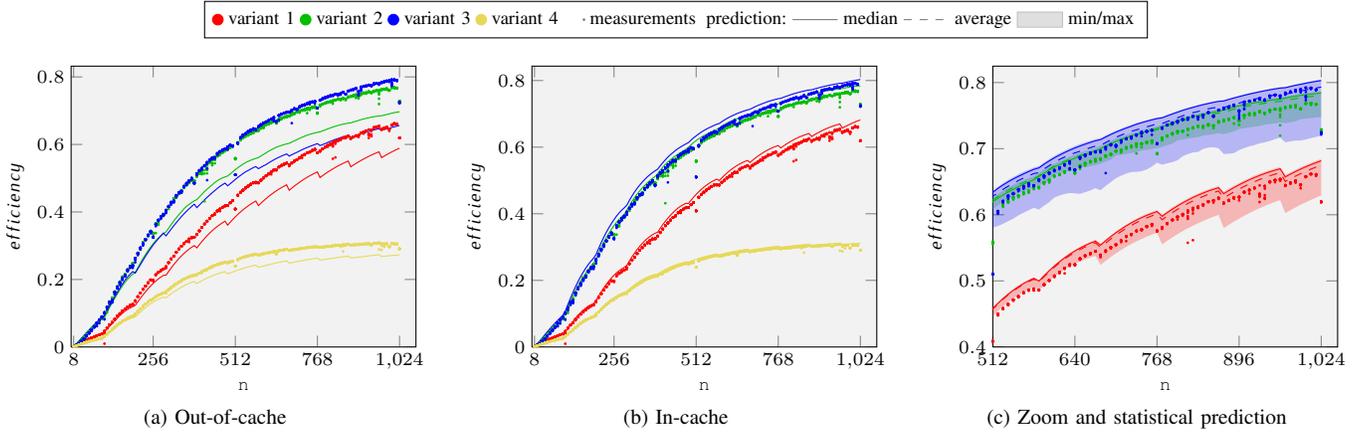
\begin{figure*}[t]
    \scriptsize
    \centering
    
    \ref*{fig:ranking.trinv.estimates.legend}

    \tikzset{external/export=true}
    \subfloat[Out-of-cache]{
        \label{fig:ranking.trinv.estimates:ooc}
        \begin{tikzpicture}
            \begin{axis}[
                width=.34\textwidth,
                xlabel={\texttt{n}},
                ylabel={\metric{efficiency}},
                xtick={8,256,512,...,1024},
                legend to name=fig:ranking.trinv.estimates.legend,
                legend columns=-1
            ]
                \addlegendimage{plot1, only marks} \label{fig:ranking.trinv.estimates:var1} \addlegendentry{variant 1\vphantom{gt/}}
                \addlegendimage{plot2, only marks} \label{fig:ranking.trinv.estimates:var2} \addlegendentry{variant 2\vphantom{gt/}}
                \addlegendimage{plot3, only marks} \label{fig:ranking.trinv.estimates:var3} \addlegendentry{variant 3\vphantom{gt/}}
                \addlegendimage{plot4, only marks} \label{fig:ranking.trinv.estimates:var4} \addlegendentry{variant 4\vphantom{gt/}\ \ \ \ }
                \addlegendimage{gray, mark size=.3pt, only marks} \addlegendentry{measurements\vphantom{gt/}}
                \addlegendimage{empty legend} \addlegendentry{prediction:\vphantom{gt/}}
                \addlegendimage{gray} \label{fig:ranking.trinv.estimates:median} \addlegendentry{median\vphantom{gt/}}
                \addlegendimage{gray, dashed} \label{fig:ranking.trinv.estimates:avg} \addlegendentry{average\vphantom{gt/}}
                \addlegendimage{draw=none, fill=gray, opacity=.25, area legend} \label{fig:ranking.trinv.estimates:minmax} \addlegendentry{min/max\vphantom{gt/}}

                \addplot[plot1, restrict x to domain=0:1024] file {figures/data/ranking.trinv.estimates/trinv1.eff.dat};
                \addplot[plot2, restrict x to domain=0:1024] file {figures/data/ranking.trinv.estimates/trinv2.eff.dat};
                \addplot[plot3, restrict x to domain=0:1024] file {figures/data/ranking.trinv.estimates/trinv3.eff.dat};
                \addplot[plot4, restrict x to domain=0:1024] file {figures/data/ranking.trinv.estimates/trinv4.eff.dat};
                \addplot[plot1, mark size=.3pt, only marks, restrict x to domain=0:1024] file {figures/data/ranking.trinv.measured/trinv1.eff.dat};
                \addplot[plot2, mark size=.3pt, only marks, restrict x to domain=0:1024] file {figures/data/ranking.trinv.measured/trinv2.eff.dat};
                \addplot[plot3, mark size=.3pt, only marks, restrict x to domain=0:1024] file {figures/data/ranking.trinv.measured/trinv3.eff.dat};
                \addplot[plot4, mark size=.3pt, only marks, restrict x to domain=0:1024] file {figures/data/ranking.trinv.measured/trinv4.eff.dat};
            \end{axis}
        \end{tikzpicture}
    }
    \hfill
    \subfloat[In-cache]{
        \label{fig:ranking.trinv.estimates:ic}
        \begin{tikzpicture}
            \begin{axis}[
                width=.34\textwidth,
                xlabel={\texttt{n}},
                ylabel={\metric{efficiency}},
                xtick={8,256,512,...,1024},
            ]
                \addplot[plot1, restrict x to domain=0:1024] file {figures/data/ranking.trinv.estimates/trinv1.eff.ic.dat};
                \addplot[plot2, restrict x to domain=0:1024] file {figures/data/ranking.trinv.estimates/trinv2.eff.ic.dat};
                \addplot[plot3, restrict x to domain=0:1024] file {figures/data/ranking.trinv.estimates/trinv3.eff.ic.dat};
                \addplot[plot4, restrict x to domain=0:1024] file {figures/data/ranking.trinv.estimates/trinv4.eff.ic.dat};
                \addplot[plot1, mark size=.3pt, only marks, restrict x to domain=0:1024] file {figures/data/ranking.trinv.measured/trinv1.eff.dat};
                \addplot[plot2, mark size=.3pt, only marks, restrict x to domain=0:1024] file {figures/data/ranking.trinv.measured/trinv2.eff.dat};
                \addplot[plot3, mark size=.3pt, only marks, restrict x to domain=0:1024] file {figures/data/ranking.trinv.measured/trinv3.eff.dat};
                \addplot[plot4, mark size=.3pt, only marks, restrict x to domain=0:1024] file {figures/data/ranking.trinv.measured/trinv4.eff.dat};
            \end{axis}
        \end{tikzpicture}
    }
    \hfill
    \subfloat[Zoom and statistical prediction]{
        \label{fig:ranking.trinv.estimates:ranges}
        \begin{tikzpicture}
            \begin{axis}[
                width=.34\textwidth,
                xlabel={\texttt{n}},
                ylabel={\metric{efficiency}},
                xmin=512,
                xtick={512,640,...,1024},
                ymin=.4,
            ]
                \addplot[opacity=.25, draw=none, fill=plot1, restrict x to domain=0:1024] file {figures/data/ranking.trinv.estimates/trinv1.eff.ranges.dat} \closedcycle;
                \addplot[opacity=.25, draw=none, fill=plot2, restrict x to domain=0:1024] file {figures/data/ranking.trinv.estimates/trinv2.eff.ranges.dat} \closedcycle;
                \addplot[opacity=.25, draw=none, fill=plot3, restrict x to domain=0:1024] file {figures/data/ranking.trinv.estimates/trinv3.eff.ranges.dat} \closedcycle;
                \addplot[opacity=.25, draw=none, fill=plot4, restrict x to domain=0:1024] file {figures/data/ranking.trinv.estimates/trinv4.eff.ranges.dat} \closedcycle;
                \addplot[plot1, restrict x to domain=0:1024] file {figures/data/ranking.trinv.estimates/trinv1.eff.ic.dat};
                \addplot[plot2, restrict x to domain=0:1024] file {figures/data/ranking.trinv.estimates/trinv2.eff.ic.dat};
                \addplot[plot3, restrict x to domain=0:1024] file {figures/data/ranking.trinv.estimates/trinv3.eff.ic.dat};
                \addplot[plot4, restrict x to domain=0:1024] file {figures/data/ranking.trinv.estimates/trinv4.eff.ic.dat};
                \addplot[plot1, dashed, restrict x to domain=0:1024] file {figures/data/ranking.trinv.estimates/trinv1.eff.avg.dat};
                \addplot[plot2, dashed, restrict x to domain=0:1024] file {figures/data/ranking.trinv.estimates/trinv2.eff.avg.dat};
                \addplot[plot3, dashed, restrict x to domain=0:1024] file {figures/data/ranking.trinv.estimates/trinv3.eff.avg.dat};
                \addplot[plot4, dashed, restrict x to domain=0:1024] file {figures/data/ranking.trinv.estimates/trinv4.eff.avg.dat};
                \addplot[plot1, mark size=.3pt, only marks, restrict x to domain=0:1024] file {figures/data/ranking.trinv.measured/trinv1.eff.dat};
                \addplot[plot2, mark size=.3pt, only marks, restrict x to domain=0:1024] file {figures/data/ranking.trinv.measured/trinv2.eff.dat};
                \addplot[plot3, mark size=.3pt, only marks, restrict x to domain=0:1024] file {figures/data/ranking.trinv.measured/trinv3.eff.dat};
                \addplot[plot4, mark size=.3pt, only marks, restrict x to domain=0:1024] file {figures/data/ranking.trinv.measured/trinv4.eff.dat};
            \end{axis}
        \end{tikzpicture}
    }

    \caption{\texttt{trinv}: Performance predictions vs. observations.}
    \label{fig:ranking.trinv.estimates}
    \tikzset{external/export=false}
\end{figure*}

We consider four blocked algorithms for the inversion of a triangular matrix.
All these algorithms partition $L$ into $6$ submatrices as
$$
    L = 
    \left(\begin{array}{ccc}
    L_{00}  &0      &0\\
    L_{10}  &L_{11} &0\\
    L_{20}  &L_{21} &L_{22}\\
    \end{array}\right).
$$
The central matrix $L_{11}$ is of size $b \times b$ (the block-size); the size of the matrix
$L_{00}$ is initially $0 \times 0$, and as the algorithm unfolds, it
increases in steps of size $b$, until $L_{00}$ spans the whole matrix $L$;
the size of $L_{22}$ decreases
accordingly; similarly, the sizes of the offdiagonal matrices are entirely determined by those of $L_{00}$. 
At each step of this matrix traversal, a sequence of update
statements is performed on the submatrices, such that $L_{00}$ contains a 
fully computed portion of $L^{-1}$. 
Once $L_{00}$ spans all of $L$, $L^{-1}$ has been computed in place.

The four algorithmic variants presented here differ in their update statements:
\begin{center}
    \begin{tabular}{|>{\cellcolor{graybg}}p{.2\textwidth}|}
        \hline
        \multicolumn{1}{|>{\cellcolor{graybg}}c|}{Variant 1} \\
        \hline
        \small
        $L_{10} \leftarrow L_{10} L_{00}$ \\
        \small
        $L_{10} \leftarrow -L_{11}^{-1} L_{10}$ \\
        \small
        $L_{11} \leftarrow L_{11}^{-1}$ \\
        \hline
    \end{tabular}
    \hspace{.25cm}
    \begin{tabular}{|>{\cellcolor{graybg}}p{.2\textwidth}|}
        \hline
        \multicolumn{1}{|>{\cellcolor{graybg}}c|}{Variant 2} \\
        \hline
        \small
        $L_{21} \leftarrow L_{22}^{-1} L_{21}$ \\
        \small
        $L_{21} \leftarrow -L_{21} L_{11}^{-1}$ \\
        \small
        $L_{11} \leftarrow L_{11}^{-1}$ \\
        \hline
    \end{tabular}

    \vspace{.25cm}

    \begin{tabular}{|>{\cellcolor{graybg}}p{.2\textwidth}|}
        \hline
        \multicolumn{1}{|>{\cellcolor{graybg}}c|}{Variant 3} \\
        \hline
        \small
        $L_{21} \leftarrow -L_{21} L_{11}^{-1}$\\
        \small
        $L_{20} \leftarrow L_{21} L_{10} + L_{20}$\\
        \small
        $L_{10} \leftarrow L_{11}^{-1} L_{10}$\\
        \small
        $L_{11} \leftarrow L_{11}^{-1}$ \\
        \hline
    \end{tabular}
    \hspace{.25cm}
    \begin{tabular}{|>{\cellcolor{graybg}}p{.2\textwidth}|}
        \hline
        \multicolumn{1}{|>{\cellcolor{graybg}}c|}{Variant 4} \\
        \hline
        \small
        $L_{21} \leftarrow -L_{22}^{-1} L_{21}$ \\
        \small
        $L_{20} \leftarrow -L_{21} L_{10} + L_{20}$ \\
        \small
        $L_{10} \leftarrow L_{10} L_{00}$ \\
        \small
        $L_{11} \leftarrow L_{11}^{-1}$ \\
        \hline
    \end{tabular}
\end{center}
They are built on top of the BLAS routines \texttt{dgemm}, \texttt{dtrsm}, and \texttt{dtrmm}; the last statement in each algorithm is a recursive call to an unblocked version of the same algorithm.
They have the following signatures:
\texttt{trinv\textit i(n, L, ldL, blocksize)}.
We consider their performance with the arguments
\begin{center}
    \small
    \texttt{%
        trinv\textit i(%
        \overparameq{n}{\textit n},
        \overparameq{L}{$L$},
        \overparameq{ldL}{\textit n},
        \overparameq{blocksize}{96})%
    },
\end{center}
varying the matrix size $\mathtt n \in \{8, 16, \ldots, 1024\}$. We consider
the performance metric \metric{efficiency}, which is computed from
\metric{ticks} as follows:
$$
    \metric{efficiency} = \frac{\frac16 \mathtt n^3 + \frac12 \mathtt n^2 + \frac13 \mathtt n}{2 \cdot \metric{ticks}}.
$$

For our performance prediction, we use performance models for \texttt{dtrsm},
\texttt{dtrmm}, \texttt{dgemm}, and the unblocked versions of the blocked
algorithms\footnote{%
    Since the unblocked versions are only invoked on small matrices, their
    models are limited to values of \texttt n below $256$.
}. The models are generated by the Modeler, with the configuration selected in
\autoref{sec:modeling.res.comp}. For each algorithm execution, we
consider the list of subroutine invocations, consisting of calls to these
routines. For instance, 
the execution of variant 1 on a matrix of size 250
with block-size $100$ produces the following invocations:\\
{ \footnotesize \ttfamily%
    dtrmm(R, L, N, N, 100, 0, 1, $L_{00}$, 250, $L_{10}$, 250)    \\
    dtrsm(L, L, N, N, 100, 0, -1, $L_{11}$, 250, $L_{10}$, 250)   \\
    trinv1(100, $L_{11}$, 250, 1)                                 \\
    dtrmm(R, L, N, N, 100, 100, 1, $L_{00}$, 250, $L_{10}$, 250)  \\
    dtrsm(L, L, N, N, 100, 100, -1, $L_{11}$, 250, $L_{10}$, 250) \\
    trinv1(100, $L_{11}$, 250, 1)                                 \\
    dtrmm(R, L, N, N, 50, 200, 1, $L_{00}$, 250, $L_{10}$, 250)   \\
    dtrsm(L, L, N, N, 50, 200, -1, $L_{11}$, 250, $L_{10}$, 250)  \\
    trinv1(50, $L_{11}$, 250, 1).                                  
}

Each invocation corresponds to the evaluation of the corresponding performance model;
the results are then accumulated, thus generating a performance prediction.

            \subsubsection{Matrix size}
            \label{sec:ranking.trinv.n}
            \autoref{fig:ranking.trinv.estimates} contains the predictions
for the four algorithm, with varying matrix size.
The left and middle panels refer to 
in-cache (\ref{fig:ranking.trinv.estimates:ic}) and
out-of-cache (\ref{fig:ranking.trinv.estimates:ooc})
scenarios, respectively. 
Since the memory locality of an actual execution is somewhere 
in between these two scenarios,
neither of the predictions matches the measurements perfectly: in-cache
overestimates the {efficiency} of the algorithms, while out-of cache
underestimates it. 
At this moment we do not yet attempt the construction of models matching the 
exact memory locality scenario of each algorithm; therefore 
in the following we use
the upper bound on {efficiency} resulting from the in-cache models.
This prediction ranks exactly all variants for all problem sizes.

The previous discussion referred to predictions for the median of the performance. In
\autoref{fig:ranking.trinv.estimates:ranges} we instead look at average,
minimum, and maximum efficiency; in order to visualize the interesting features, we
only present the top right portion of the graph ($n \geq 512$ and $0.5 \leq
\metric{efficiency} \leq 0.8$). The ranges between minimum and maximum
(\ref{fig:ranking.trinv.estimates:minmax}) cover almost all the observations 
of the corresponding algorithms, giving a good idea of the expected results;
their height is due to the presence of outliers.

The average
(\ref{fig:ranking.trinv.estimates:avg}) is closer to the measured algorithm
performance than the previously used median. Nevertheless, relying on the
average predictions, is dangerous, since they are obtained for models generated
with an error bound on the median and are influenced by outliers.

Altogether, we predicted performance for varying matrix sizes with highly satisfactory results.

            \subsubsection{Block-size}
            \label{sec:ranking.trinv.b}
            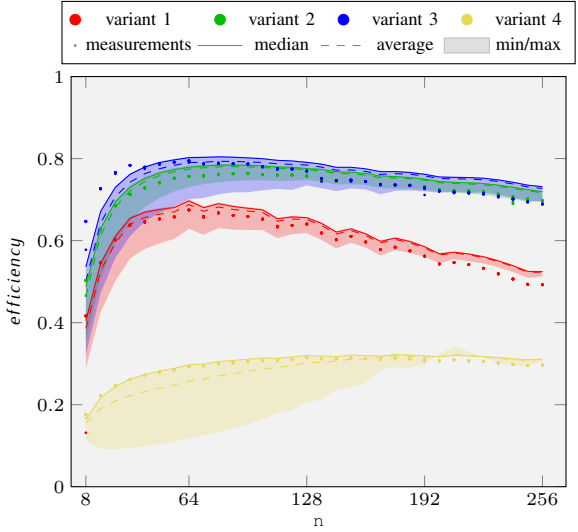
\begin{figure}[t]
    \scriptsize
    \centering
    \tikzset{external/export=true}

    \begin{tikzpicture}
        \begin{axis}[
            onecolplot,
            xlabel={\texttt{n}},
            ylabel={\metric{efficiency}},
            ymin=0,
            ymax=1,
            xtick={8,64,128,...,256},
            legend columns=4,
            legend style={
                at={(0.5,1.03)},
                anchor=south
            }
        ]
            \addlegendimage{plot1, only marks} \label{fig:ranking.trinv.b:var1} \addlegendentry{variant 1\vphantom{gt/}}
            \addlegendimage{plot2, only marks} \label{fig:ranking.trinv.b:var2} \addlegendentry{variant 2\vphantom{gt/}}
            \addlegendimage{plot3, only marks} \label{fig:ranking.trinv.b:var3} \addlegendentry{variant 3\vphantom{gt/}}
            \addlegendimage{plot4, only marks} \label{fig:ranking.trinv.b:var4} \addlegendentry{variant 4\vphantom{gt/}}
            \addlegendimage{gray, mark size=.3pt, only marks} \addlegendentry{measurements\vphantom{gt/}}
            \addlegendimage{gray} \label{fig:ranking.trinv.b:median} \addlegendentry{median\vphantom{gt/}}
            \addlegendimage{gray, dashed} \label{fig:ranking.trinv.b:avg} \addlegendentry{average\vphantom{gt/}}
            \addlegendimage{draw=none, fill=gray, opacity=.25, area legend} \label{fig:ranking.trinv.b:minmax} \addlegendentry{min/max\vphantom{gt/}}

            \addplot[opacity=.25, draw=none, fill=plot1, restrict x to domain=0:256] file {figures/data/ranking.trinv.b/trinv1.eff.minmax.dat} \closedcycle;
            \addplot[opacity=.25, draw=none, fill=plot2, restrict x to domain=0:256] file {figures/data/ranking.trinv.b/trinv2.eff.minmax.dat} \closedcycle;
            \addplot[opacity=.25, draw=none, fill=plot3, restrict x to domain=0:256] file {figures/data/ranking.trinv.b/trinv3.eff.minmax.dat} \closedcycle;
            \addplot[opacity=.25, draw=none, fill=plot4, restrict x to domain=0:256] file {figures/data/ranking.trinv.b/trinv4.eff.minmax.dat} \closedcycle;
            \addplot[plot1, restrict x to domain=0:256] file {figures/data/ranking.trinv.b/trinv1.eff.med.dat};
            \addplot[plot2, restrict x to domain=0:256] file {figures/data/ranking.trinv.b/trinv2.eff.med.dat};
            \addplot[plot3, restrict x to domain=0:256] file {figures/data/ranking.trinv.b/trinv3.eff.med.dat};
            \addplot[plot4, restrict x to domain=0:256] file {figures/data/ranking.trinv.b/trinv4.eff.med.dat};
            \addplot[plot1, dashed, restrict x to domain=0:256] file {figures/data/ranking.trinv.b/trinv1.eff.avg.dat};
            \addplot[plot2, dashed, restrict x to domain=0:256] file {figures/data/ranking.trinv.b/trinv2.eff.avg.dat};
            \addplot[plot3, dashed, restrict x to domain=0:256] file {figures/data/ranking.trinv.b/trinv3.eff.avg.dat};
            \addplot[plot4, dashed, restrict x to domain=0:256] file {figures/data/ranking.trinv.b/trinv4.eff.avg.dat};
            \addplot[plot1, mark size=.3pt, only marks, restrict x to domain=0:256] file {figures/data/ranking.trinv.b/trinv1.eff.meas.dat};
            \addplot[plot2, mark size=.3pt, only marks, restrict x to domain=0:256] file {figures/data/ranking.trinv.b/trinv2.eff.meas.dat};
            \addplot[plot3, mark size=.3pt, only marks, restrict x to domain=0:256] file {figures/data/ranking.trinv.b/trinv3.eff.meas.dat};
            \addplot[plot4, mark size=.3pt, only marks, restrict x to domain=0:256] file {figures/data/ranking.trinv.b/trinv4.eff.meas.dat};
        \end{axis}
    \end{tikzpicture}

    \caption{Block-size optimization for \texttt{trinv}.}
    \label{fig:ranking.trinv.b:eff}
    \tikzset{external/export=false}
\end{figure}

We now turn our attention to our second point of interest: 
tuning the block-size to yield the best efficiency.
For this purpose, we fix the matrix size to $\mathtt n = 1000$ and vary the block-size:
\begin{center}
    \small
    \texttt{%
        trinv\textit i(%
        \overparameq{n}{1000},
        \overparameq{L}{$L$},
        \overparameq{ldL}{1000},
        \overparameq{blocksize}{\textit{blocksize}})%
    }.
\end{center}
The resulting predictions (\autoref{fig:ranking.trinv.b:eff}) capture very well
the behavior for the most efficient block-sizes (between $48$ and
$128$). For instance, Variant~3~(\ref{fig:ranking.trinv.b:var3}) ---the fastest--- attains
its top performance with a block-size of 64 according to both the measurements
and our prediction.

The quality of our prediction decreases for very large and very small
block-sizes. For our goal ---determining the fastest algorithm configuration--- 
the low accuracy in regions with low performance is not an issue.
The decreasing accuracy in the maximum and average performance for
variant~4~(\ref{fig:ranking.trinv.b:var4}), resulting from measurement outliers
in the model generation, shows that these quantities are less well suited for
predictions. The median on the other hand is very reliable.

            \diff{
            \subsubsection{Sandy Bridge}
            \label{sec:ranking.sandy}
            \begin{figure}[t]
    \scriptsize
    \centering
    \tikzset{external/export=true}

    \begin{tikzpicture}
        \begin{axis}[
            onecolplot,
            xlabel={\texttt{n}},
            ylabel={\metric{efficiency}},
            xmin=512,
            ymin=.4,
            xtick={8,256,512,...,1024},
            legend columns=4,
            legend style={
                at={(0.5,1.03)},
                anchor=south
            }
        ]
            \addlegendimage{plot1, only marks} \label{fig:ranking.trinv.sandy:var1} \addlegendentry{variant 1\vphantom{gt/}}
            \addlegendimage{plot2, only marks} \label{fig:ranking.trinv.sandy:var2} \addlegendentry{variant 2\vphantom{gt/}}
            \addlegendimage{plot3, only marks} \label{fig:ranking.trinv.sandy:var3} \addlegendentry{variant 3\vphantom{gt/}}
            \addlegendimage{plot4, only marks} \label{fig:ranking.trinv.sandy:var4} \addlegendentry{variant 4\vphantom{gt/}}
            \addlegendimage{gray, mark size=.3pt, only marks} \addlegendentry{measurements\vphantom{gt/}}
            \addlegendimage{gray} \label{fig:ranking.trinv.sandy:median} \addlegendentry{median\vphantom{gt/}}
            \addlegendimage{gray, dashed} \label{fig:ranking.trinv.sandy:avg} \addlegendentry{average\vphantom{gt/}}
            \addlegendimage{draw=none, fill=gray, opacity=.25, area legend} \label{fig:ranking.trinv.sandy:minmax} \addlegendentry{min/max\vphantom{gt/}}

            \addplot[opacity=.25, draw=none, fill=plot1, restrict x to domain=0:1024] file {figures/data/ranking.trinv.sandy/trinv1.eff.minmax.dat} \closedcycle;
            \addplot[opacity=.25, draw=none, fill=plot2, restrict x to domain=0:1024] file {figures/data/ranking.trinv.sandy/trinv2.eff.minmax.dat} \closedcycle;
            \addplot[opacity=.25, draw=none, fill=plot3, restrict x to domain=0:1024] file {figures/data/ranking.trinv.sandy/trinv3.eff.minmax.dat} \closedcycle;
            \addplot[opacity=.25, draw=none, fill=plot4, restrict x to domain=0:1024] file {figures/data/ranking.trinv.sandy/trinv4.eff.minmax.dat} \closedcycle;
            \addplot[plot1, restrict x to domain=0:1024] file {figures/data/ranking.trinv.sandy/trinv1.eff.med.dat};
            \addplot[plot2, restrict x to domain=0:1024] file {figures/data/ranking.trinv.sandy/trinv2.eff.med.dat};
            \addplot[plot3, restrict x to domain=0:1024] file {figures/data/ranking.trinv.sandy/trinv3.eff.med.dat};
            \addplot[plot4, restrict x to domain=0:1024] file {figures/data/ranking.trinv.sandy/trinv4.eff.med.dat};
            \addplot[plot1, dashed, restrict x to domain=0:1024] file {figures/data/ranking.trinv.sandy/trinv1.eff.avg.dat};
            \addplot[plot2, dashed, restrict x to domain=0:1024] file {figures/data/ranking.trinv.sandy/trinv2.eff.avg.dat};
            \addplot[plot3, dashed, restrict x to domain=0:1024] file {figures/data/ranking.trinv.sandy/trinv3.eff.avg.dat};
            \addplot[plot4, dashed, restrict x to domain=0:1024] file {figures/data/ranking.trinv.sandy/trinv4.eff.avg.dat};
            \addplot[plot1, mark size=.3pt, only marks, restrict x to domain=0:1024] file {figures/data/ranking.trinv.sandy/trinv1.eff.meas.dat};
            \addplot[plot2, mark size=.3pt, only marks, restrict x to domain=0:1024] file {figures/data/ranking.trinv.sandy/trinv2.eff.meas.dat};
            \addplot[plot3, mark size=.3pt, only marks, restrict x to domain=0:1024] file {figures/data/ranking.trinv.sandy/trinv3.eff.meas.dat};
            \addplot[plot4, mark size=.3pt, only marks, restrict x to domain=0:1024] file {figures/data/ranking.trinv.sandy/trinv4.eff.meas.dat};
        \end{axis}
    \end{tikzpicture}

    \caption{\texttt{trinv}: Predictions and observations on Sandy Bridge.}
    \label{fig:ranking.trinv.sandy:eff}
    \tikzset{external/export=false}
\end{figure}
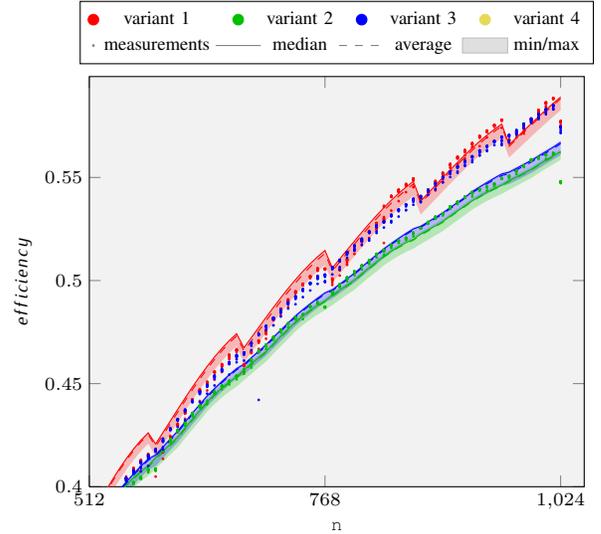

We now move to a newer CPU architecture: Intel's Sandy Bridge-EP E5-2670 running at 2.60GHz.
On this system, we use the OpenBLAS library, for which we generate a new set of performance models.
The predictions for \texttt{trinv} obtained from these models are shown in \autoref{fig:ranking.trinv.sandy:eff}.
As for the Harpertown, the least efficient variant is \#4~(\ref{fig:ranking.trinv.sandy:var4}); 
however, on this system the fastest variant is \#1~(\ref{fig:ranking.trinv.sandy:var1}).
Besides, while on this processor the predictions for some variants are not as
accurate as on the Harpertown, the ranking is still carried out correctly.

            }

            \subsubsection{Shared memory parallelism}
            \label{sec:ranking.trinv.parallel}
            Our modeling strategy applies well to shared memory architectures too.
In this next experiment, we utilize all the 8 cores of the \diff{Sandy Bridge} processor.
We achieve parallelism by linking the routines for 
matrix inversions with the multithreaded version of the \diff{OpenBLAS} library. 
The performance predictions are obtained from performance models
generated from measurements of the multithreaded BLAS routines.

\diff{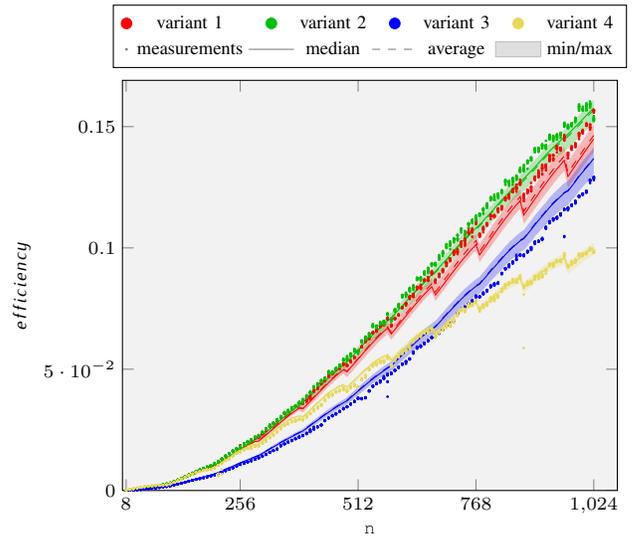
\begin{figure}[t]
    \scriptsize
    \centering
    \tikzset{external/export=true}

    \begin{tikzpicture}
        \begin{axis}[
            onecolplot,
            xlabel={\texttt{n}},
            ylabel={\metric{efficiency}},
            xtick={8,256,512,...,1024},
            legend columns=4,
            legend style={
                at={(0.5,1.03)},
                anchor=south
            }
        ]
            \addlegendimage{plot1, only marks} \addlegendentry{variant 1\vphantom{gt/}}
            \label{fig:ranking.trinv.parallel:var1}
            \addlegendimage{plot2, only marks} \addlegendentry{variant 2\vphantom{gt/}}
            \label{fig:ranking.trinv.parallel:var2}
            \addlegendimage{plot3, only marks} \addlegendentry{variant 3\vphantom{gt/}}
            \label{fig:ranking.trinv.parallel:var3}
            \addlegendimage{plot4, only marks} \addlegendentry{variant 4\vphantom{gt/}}
            \label{fig:ranking.trinv.parallel:var4}
            \addlegendimage{gray, mark size=.3pt, only marks} \addlegendentry{measurements\vphantom{gt/}}
            \addlegendimage{gray} \addlegendentry{median\vphantom{gt/}}
            \addlegendimage{gray, dashed} \addlegendentry{average\vphantom{gt/}}
            \addlegendimage{draw=none, fill=gray, opacity=.25, area legend} \addlegendentry{min/max\vphantom{gt/}}

            \addplot[opacity=.25, draw=none, fill=plot1, restrict x to domain=0:1024] file {figures/data/ranking.trinv.parallel.model/trinv1.eff.ranges.dat} \closedcycle;
            \addplot[opacity=.25, draw=none, fill=plot2, restrict x to domain=0:1024] file {figures/data/ranking.trinv.parallel.model/trinv2.eff.ranges.dat} \closedcycle;
            \addplot[opacity=.25, draw=none, fill=plot3, restrict x to domain=0:1024] file {figures/data/ranking.trinv.parallel.model/trinv3.eff.ranges.dat} \closedcycle;
            \addplot[opacity=.25, draw=none, fill=plot4, restrict x to domain=0:1024] file {figures/data/ranking.trinv.parallel.model/trinv4.eff.ranges.dat} \closedcycle;
            \addplot[plot1, restrict x to domain=0:1024] file {figures/data/ranking.trinv.parallel.model/trinv1.eff.ic.dat};
            \addplot[plot2, restrict x to domain=0:1024] file {figures/data/ranking.trinv.parallel.model/trinv2.eff.ic.dat};
            \addplot[plot3, restrict x to domain=0:1024] file {figures/data/ranking.trinv.parallel.model/trinv3.eff.ic.dat};
            \addplot[plot4, restrict x to domain=0:1024] file {figures/data/ranking.trinv.parallel.model/trinv4.eff.ic.dat};
            \addplot[plot1, dashed, restrict x to domain=0:1024] file {figures/data/ranking.trinv.parallel.model/trinv1.eff.avg.dat};
            \addplot[plot2, dashed, restrict x to domain=0:1024] file {figures/data/ranking.trinv.parallel.model/trinv2.eff.avg.dat};
            \addplot[plot3, dashed, restrict x to domain=0:1024] file {figures/data/ranking.trinv.parallel.model/trinv3.eff.avg.dat};
            \addplot[plot4, dashed, restrict x to domain=0:1024] file {figures/data/ranking.trinv.parallel.model/trinv4.eff.avg.dat};

            \addplot[plot1, mark size=.3pt, only marks, restrict x to domain=0:1024] file {figures/data/ranking.trinv.parallel/trinv1.eff.dat};
            \addplot[plot2, mark size=.3pt, only marks, restrict x to domain=0:1024] file {figures/data/ranking.trinv.parallel/trinv2.eff.dat};
            \addplot[plot3, mark size=.3pt, only marks, restrict x to domain=0:1024] file {figures/data/ranking.trinv.parallel/trinv3.eff.dat};
            \addplot[plot4, mark size=.3pt, only marks, restrict x to domain=0:1024] file {figures/data/ranking.trinv.parallel/trinv4.eff.dat};
        \end{axis}
    \end{tikzpicture}

    \caption{\texttt{trinv}: Predictions and observations on 8 cores.}
    \label{fig:ranking.trinv.parallel}
    \tikzset{external/export=false}
\end{figure}}

The resulting predictions along with measurements of \texttt{trinv}'s
performance are shown in \autoref{fig:ranking.trinv.parallel}.
\diff{The predictions match the experiments very closely and allow us to rank the algorithmic variants correctly for all problem sizes.
Moreover,
the multithreaded variants exhibit two interesting features:
At about $\mathtt n = 650$, variants~3~(\ref{fig:ranking.trinv.parallel:var3}) and 4~(\ref{fig:ranking.trinv.parallel:var4}) crossover, and in contrast to the sequential case,
variants~1~(\ref{fig:ranking.trinv.parallel:var1}) and 2~(\ref{fig:ranking.trinv.parallel:var2}) are faster than variant 3~(\ref{fig:ranking.trinv.parallel:var1}).
Both these features also appear in our predictions.
}

        \subsection{Sylvester Equation: Solving \texorpdfstring{$L X + X U = C$ for $X$}{L X + X U = C for X}}
        \label{sec:ranking.sylv}
        We now study of a more complicated operation: the solution of the Sylvester
equation. This operation, encountered in control theory, is generally of the
form $A X + X B = C$, where $A \in \mathbb R^{m \times m}$, $B \in \mathbb R^{n
\times n}$, and $C \in \mathbb R^{m \times n}$ are given, and $X \in \mathbb
R^{m \times n}$ is to be computed. We consider a special case, where $A$ and
$B$ are lower and upper triangular matrices, respectively: $L X + X U = C$.

With {\sc{Cl\makebox[.58\width][c]{1}ck}} \cite{diego1, diego2}, a tool for the
automatic generation of blocked algorithms, we generated code for 16 algorithmic variants. 
Each of them takes as input the three matrices $L$, $U$, and $X$; $X$
initially contains the input matrix $C$ and is overwritten with the solution 
to the equation.

The exemplary update statements for variants 1 and 16 are given below. There, 
$\Omega(L, U, X_{ij})$ denotes a recursive invocation to the Sylvester
equations solver  for the smaller matrix $X_{ij}$. The signature of this solver is
\texttt{sylv\textit i(m, n, L, ldL, U, ldU, X, ldX, blocksize)} with
$\texttt{\textit i} \in \{1, \ldots, 16\}$.

\begin{center}
    \newcommand{\sylv}{\Omega}
    \begin{tabular}{|>{\cellcolor{graybg}}p{.2\textwidth}|}
        \hline
        \multicolumn{1}{|>{\cellcolor{graybg}}c|}{Variant 1} \\
        \hline
        \small
         $X_{01} \leftarrow X_{01} - X_{00} U_{01}$ \\
        \small
         $X_{10} \leftarrow X_{10} - L_{10} X_{00}$ \\
        \small
         $X_{01} \leftarrow \sylv(L_{00}, U_{11}, X_{01})$ \\
        \small
         $X_{10} \leftarrow \sylv(L_{11}, U_{00}, X_{10})$ \\
        \small
         $X_{11} \leftarrow X_{11} - X_{10} U_{01}$ \\
        \small
         $X_{11} \leftarrow X_{11} - L_{10} X_{01}$ \\
        \small
         $X_{11} \leftarrow \sylv(L_{11}, U_{11}, X_{11})$ \\
        \hline
    \end{tabular}
    \hspace{.25cm}
    \begin{tabular}{|>{\cellcolor{graybg}}p{.2\textwidth}|}
        \hline
        \multicolumn{1}{|>{\cellcolor{graybg}}c|}{Variant 16} \\
        \hline
        \small
         $X_{11} \leftarrow \sylv(L_{11}, U_{11}, X_{11})$ \\
        \small
         $X_{12} \leftarrow X_{12} - X_{11} U_{12}$ \\
        \small
         $X_{21} \leftarrow X_{21} - L_{21} X_{11}$ \\
        \small
         $X_{12} \leftarrow \sylv(L_{11}, U_{22}, X_{12})$ \\
        \small
         $X_{21} \leftarrow \sylv(L_{22}, U_{11}, X_{21})$ \\
        \small
         $X_{22} \leftarrow X_{22} - X_{21} U_{12}$ \\
        \small
         $X_{22} \leftarrow X_{22} - L_{21} X_{12}$ \\
        \hline
    \end{tabular}
\end{center}

These blocked algorithms differ 
from those for \texttt{trinv\textit i} in 
in a number of ways. 
\begin{itemize}
    \item They operate on three matrices, overwriting one of them with the
    output.

    \item The input matrices are of different sizes, and not all of
     them are square: $L \in \mathbb R^{\mathtt m \times \mathtt m}$,
     $U \in \mathbb R^{\mathtt n \times \mathtt n}$, and
     $X \in \mathbb R^{\mathtt m \times \mathtt n}$. Moreover, the matrices are
     traversed along the diagonal as far as possible and then along
     the remaining dimension.

    \item At each iteration, the algorithms perform three recursive
          calls to $\Omega$. These operate not only on the
          $X_{11} \in \mathbb
          R^{\mathtt{blocksize} \times \mathtt{blocksize}}$ but also
          on the matrix panels $X_{01}$, $X_{10}$, $X_{12}$, and
          $X_{21}$. For the latter, our C implementation invokes the
          blocked algorithms recursively; only the small matrices
          $X_{11}$ trigger their unblocked versions.
\end{itemize}

\diff{\begin{figure}[t]
    \scriptsize
    \centering
    \tikzset{external/export=true}
    
    \begin{tikzpicture}
        \begin{axis}[
            onecolplot,
            xlabel={\texttt{n}},
            ylabel={\metric{efficiency}},
            xtick={8,256,512,...,1024},
            legend columns=8,
            legend style={
                at={(0.5,1.03)},
                anchor=south
            }
        ]
            \addlegendimage{plot1, mark=*} \label{fig:ranking.sylv.n:var1} \addlegendentry{1}
            \addlegendimage{plot2, mark=*} \label{fig:ranking.sylv.n:var2} \addlegendentry{2}
            \addlegendimage{plot3, mark=*} \label{fig:ranking.sylv.n:var3} \addlegendentry{3}
            \addlegendimage{plot4, mark=*} \label{fig:ranking.sylv.n:var4} \addlegendentry{4}
            \addlegendimage{plot5, mark=*} \label{fig:ranking.sylv.n:var5} \addlegendentry{5}
            \addlegendimage{plot6, mark=*} \label{fig:ranking.sylv.n:var6} \addlegendentry{6}
            \addlegendimage{plot7, mark=*} \label{fig:ranking.sylv.n:var7} \addlegendentry{7}
            \addlegendimage{plot8, mark=*} \label{fig:ranking.sylv.n:var8} \addlegendentry{8}
            \addlegendimage{plot1, mark=x, dashed} \label{fig:ranking.sylv.n:var9} \addlegendentry{9}
            \addlegendimage{plot2, mark=x, dashed} \label{fig:ranking.sylv.n:var10} \addlegendentry{10}
            \addlegendimage{plot3, mark=x, dashed} \label{fig:ranking.sylv.n:var11} \addlegendentry{11}
            \addlegendimage{plot4, mark=x, dashed} \label{fig:ranking.sylv.n:var12} \addlegendentry{12}
            \addlegendimage{plot5, mark=x, dashed} \label{fig:ranking.sylv.n:var13} \addlegendentry{13}
            \addlegendimage{plot6, mark=x, dashed} \label{fig:ranking.sylv.n:var14} \addlegendentry{14}
            \addlegendimage{plot7, mark=x, dashed} \label{fig:ranking.sylv.n:var15} \addlegendentry{15}
            \addlegendimage{plot8, mark=x, dashed} \label{fig:ranking.sylv.n:var16} \addlegendentry{16}

            \addplot[plot1, mark size=.3pt, only marks, restrict x to domain=0:1024] file {figures/data/ranking.sylv.n/sylv1.eff.meas.dat};
            \addplot[plot2, mark size=.3pt, only marks, restrict x to domain=0:1024] file {figures/data/ranking.sylv.n/sylv2.eff.meas.dat};
            \addplot[plot3, mark size=.3pt, only marks, restrict x to domain=0:1024] file {figures/data/ranking.sylv.n/sylv3.eff.meas.dat};
            \addplot[plot4, mark size=.3pt, only marks, restrict x to domain=0:1024] file {figures/data/ranking.sylv.n/sylv4.eff.meas.dat};
            \addplot[plot5, mark size=.3pt, only marks, restrict x to domain=0:1024] file {figures/data/ranking.sylv.n/sylv5.eff.meas.dat};
            \addplot[plot6, mark size=.3pt, only marks, restrict x to domain=0:1024] file {figures/data/ranking.sylv.n/sylv6.eff.meas.dat};
            \addplot[plot7, mark size=.3pt, only marks, restrict x to domain=0:1024] file {figures/data/ranking.sylv.n/sylv7.eff.meas.dat};
            \addplot[plot8, mark size=.3pt, only marks, restrict x to domain=0:1024] file {figures/data/ranking.sylv.n/sylv8.eff.meas.dat};
            \addplot[plot1, mark size=.3pt, only marks, mark=x, restrict x to domain=0:1024] file {figures/data/ranking.sylv.n/sylv9.eff.meas.dat};
            \addplot[plot2, mark size=.3pt, only marks, mark=x, restrict x to domain=0:1024] file {figures/data/ranking.sylv.n/sylv10.eff.meas.dat};
            \addplot[plot3, mark size=.3pt, only marks, mark=x, restrict x to domain=0:1024] file {figures/data/ranking.sylv.n/sylv11.eff.meas.dat};
            \addplot[plot4, mark size=.3pt, only marks, mark=x, restrict x to domain=0:1024] file {figures/data/ranking.sylv.n/sylv12.eff.meas.dat};
            \addplot[plot5, mark size=.3pt, only marks, mark=x, restrict x to domain=0:1024] file {figures/data/ranking.sylv.n/sylv13.eff.meas.dat};
            \addplot[plot6, mark size=.3pt, only marks, mark=x, restrict x to domain=0:1024] file {figures/data/ranking.sylv.n/sylv14.eff.meas.dat};
            \addplot[plot7, mark size=.3pt, only marks, mark=x, restrict x to domain=0:1024] file {figures/data/ranking.sylv.n/sylv15.eff.meas.dat};
            \addplot[plot8, mark size=.3pt, only marks, mark=x, restrict x to domain=0:1024] file {figures/data/ranking.sylv.n/sylv16.eff.meas.dat};

            \addplot[plot1, restrict x to domain=0:1024] file {figures/data/ranking.sylv.n/sylv1.eff.med.dat};
            \addplot[plot2, restrict x to domain=0:1024] file {figures/data/ranking.sylv.n/sylv2.eff.med.dat};
            \addplot[plot3, restrict x to domain=0:1024] file {figures/data/ranking.sylv.n/sylv3.eff.med.dat};
            \addplot[plot4, restrict x to domain=0:1024] file {figures/data/ranking.sylv.n/sylv4.eff.med.dat};
            \addplot[plot5, restrict x to domain=0:1024] file {figures/data/ranking.sylv.n/sylv5.eff.med.dat};
            \addplot[plot6, restrict x to domain=0:1024] file {figures/data/ranking.sylv.n/sylv6.eff.med.dat};
            \addplot[plot7, restrict x to domain=0:1024] file {figures/data/ranking.sylv.n/sylv7.eff.med.dat};
            \addplot[plot8, restrict x to domain=0:1024] file {figures/data/ranking.sylv.n/sylv8.eff.med.dat};
            \addplot[plot1, dashed, restrict x to domain=0:1024] file {figures/data/ranking.sylv.n/sylv9.eff.med.dat};
            \addplot[plot2, dashed, restrict x to domain=0:1024] file {figures/data/ranking.sylv.n/sylv10.eff.med.dat};
            \addplot[plot3, dashed, restrict x to domain=0:1024] file {figures/data/ranking.sylv.n/sylv11.eff.med.dat};
            \addplot[plot4, dashed, restrict x to domain=0:1024] file {figures/data/ranking.sylv.n/sylv12.eff.med.dat};
            \addplot[plot5, dashed, restrict x to domain=0:1024] file {figures/data/ranking.sylv.n/sylv13.eff.med.dat};
            \addplot[plot6, dashed, restrict x to domain=0:1024] file {figures/data/ranking.sylv.n/sylv14.eff.med.dat};
            \addplot[plot7, dashed, restrict x to domain=0:1024] file {figures/data/ranking.sylv.n/sylv15.eff.med.dat};
            \addplot[plot8, dashed, restrict x to domain=0:1024] file {figures/data/ranking.sylv.n/sylv16.eff.med.dat};
        \end{axis}
    \end{tikzpicture}

    \caption{\texttt{sylv}: Predictions (---) vs. observations ($\bullet, \times$).}
    \label{fig:ranking.sylv.n:meas}
    \tikzset{external/export=false}
\end{figure}}


In our tests, we consider the case
\begin{center}
    \small
    \texttt{%
        sylv\textit i(%
        \overparameq{m}{\textit n},
        \overparameq{n}{\textit n},
        \overparameq{L}{$L$},
        \overparameq{ldL}{\textit n},
        \overparameq{U}{$U$},
        \overparameq{ldU}{\textit n},
        \overparameq{X}{$X$},
        \overparameq{ldX}{\textit n},
        \overparameq{blocksize}{96})%
    }.
\end{center}
All matrices are of size $\mathtt n \times \mathtt n$, $\mathtt n \in \{8, 16,
\ldots, 1024\}$  and we use 96 as  block-size. Figures~\ref{fig:ranking.sylv.n:meas}
compares our predictions for these algorithms with corresponding measurements
of their implementations, where
$$
    \metric{efficiency} = \frac{\mathtt n^3 + \mathtt n^2}{2 \metric{ticks}}.
$$
We observe significantly different performances across algorithms: At $\mathtt
n = 1024$ variant~1~(\ref{fig:ranking.sylv.n:var1}) is more than 20 times
faster than variant~13~(\ref{fig:ranking.sylv.n:var13}). Indeed, twelve of the variants attain a performance below 2\%, while the other four reach values around 20\%.
In such a scenario, it is first  crucial to tell apart the two groups, and then to correctly rank the four top variants. Although our individual predictions are not especially accurate, they fulfill the objective perfectly, separating the groups, and ordering variants~1~(\ref{fig:ranking.sylv.n:var1}), 2~(\ref{fig:ranking.sylv.n:var2}), 5~(\ref{fig:ranking.sylv.n:var5}), and 6~(\ref{fig:ranking.sylv.n:var6})
as the top most efficient algorithms.

    \section{Conclusion} 
    \label{sec:conclusion}
    In this article, we presented an approach to analyze and model the performance
of dense linear algebra routines. Our goal was to rank a given collection of
blocked algorithms according to their performance and to optimize their
configuration, without executing them. Towards this goal, we created a
performance modeling tool, the Modeler, that automatic generates models for
BLAS and LAPACK routines. 
We introduced two strategies to originate piecewise polynomial models,
to favor either speed or accuracy.  Upon creation, the set of models
is stored in an easily accessible repository, for easy access and
evaluation.  In order to predict the performance of a blocked
algorithm, the performance models of its building blocks are then
evaluated and combined.

We showed that the approach is applicable to operations with
numerous algorithm variants both on single- and multi-core systems; 
experiments confirmed that our
predictions are able to both correctly tell apart the variants according to their
performance, and to identify the optimal algorithmic block-size.

    \bibliography{references}
    \bibliographystyle{IEEEtran}
    
\end{document}